\documentclass[a4paper,11pt]{article}

\usepackage{jcappub} 

\usepackage[T1]{fontenc} 
\usepackage{graphicx}
\usepackage{amssymb}

\newcommand{\dxi}{\mbox{$\dot{\xi}$}}
\newcommand{\ddxi}{\mbox{$\ddot{\xi}$}}

\title{Constraints on modified Gauss-Bonnet gravity during big bang nucleosynthesis}

\author[a,b,1]{Motohiko Kusakabe,\note{Corresponding author,\\JSPS Postdoctoral Fellow for Research Abroad, \\Present address:  Department of Physics, University of Notre Dame, Notre Dame, Indiana 46556, USA}}
\author[c]{Seoktae Koh,}
\author[a]{K. S. Kim}
\author[b]{and Myung-Ki Cheoun$^{2}$}

\affiliation[a]{School of Liberal Arts and Science, Korea Aerospace University,\\ Goyang 412-791, Korea}
\affiliation[b]{Department of Physics, Soongsil University,\\ Seoul 156-743, Korea}
\affiliation[c]{Department of Science Education, Jeju National University, \\Jeju 690-756, Korea}

\emailAdd{mkusakab@nd.edu}
\emailAdd{kundol.koh@jejunu.ac.kr}
\emailAdd{kyungsik@kau.ac.kr}
\emailAdd{cheoun@ssu.ac.kr}

\abstract{
The modified gravity is considered to be one of possible explanations of the accelerated expansions of the present and the early universe.  We study effects of the modified gravity on big bang nucleosynthesis (BBN).  If effects of the modified gravity are significant during the BBN epoch, they should be observed as changes of primordial light element abundances.  We assume a $f(G)$ term with the Gauss-Bonnet term $G$, during the BBN epoch.  A power-law relation of $df/dG \propto t^p$ where $t$ is the cosmic time was assumed for the function $f(G)$ as an example case.  We solve time evolutions of physical variables during BBN in the $f(G)$ gravity model numerically, and analyzed calculated results.  It is found that a proper solution for the cosmic expansion rate can be lost in some parameter region.  In addition, we show that calculated results of primordial light element abundances can be significantly different from observational data.  Especially, observational limits on primordial D abundance leads to the strongest constraint on the $f(G)$ gravity.  We then derive constraints on parameters of the $f(G)$ gravity taking into account the existence of the solution of expansion rate and final light element abundances. 
}


\begin{document}
\maketitle
\flushbottom

\section{Introduction}\label{sec1}

The accelerated expansion of the present universe has been verified by observational data on magnitude-redshift  of type Ia supernovae \cite{Riess:1998cb,Perlmutter:1998np}.  In standard cosmology, this acceleration is described by the cosmological constant or the $\Lambda$ term in the gravitational action.
Dark energy or modified gravity theory are also widely considered to explain the late time acceleration and one of the challenges is to discriminate  the dark energy model and the modified gravity theory through the observations.  {\it For example,} each of those models shows the different  histories of cosmic expansion (e.g., \cite{Cognola:2006eg}) and growth rate for the large scale structure (e.g., \cite{Li:2007jm,DeFelice2009}).

  One of feasible models of the modified gravity includes the Gauss-Bonnet
 (GB) correction, $G \equiv R^2 -4 R_{\mu\nu} R^{\mu\nu} + R_{\mu\nu\rho\sigma} R^{\mu\nu\rho\sigma}$,
 which is a simple extension of Einstein gravity.
  The GB term does not produce any ghost particles as well as any problems with unitarity \cite{Zwiebach:1985uq}.
  Additionally, the equations of motion do not contain  higher than second order
 in temporal derivatives, so there does not exist any instability problem.
 We generalize the GB correction as $f(G)$ in which $f(G)$ is a general function of $G$.  Cosmological effects of this $f(G)$ model have been investigated intensively as a simple extension of the general relativity.  In general, modifications of the general relativity are restricted from observed celestial motions in the solar system. 
However, no correction to the Newton's law and no instability are induced at the present universe in the $f(G)$ models when the functional shape of $f(G)$ is fine-tuned \cite{Cognola:2006eg}.  Therefore, the $f(G)$ gravity can escape from the constraint from the Newton's law, so that it can be a candidate theory for the accelerated expansion of the universe.

It has been suggested that epochs realized in the standard $\Lambda$CDM model can be described by  some functions $f(G)$ in the $f(G)$ models and that additional degrees of freedom in the $f(G)$ model can be tested in future observations of cosmic expansion (e.g., \cite{Elizalde:2010jx}).  Constrains on modified gravity models are, however, derived not only from the cosmic expansion rate but also from considerations of the cosmological perturbation (e.g., \cite{Li:2007jm,DeFelice2009}).  The latter constraint has been found to be very strong.  It has been argued that it is theoretically interesting to check whether viable $f(G)$ models satisfy the weak energy condition, i.e., $T_{\mu \nu} U^\mu U^\nu \geq 0$ for all timelike vectors $U^\mu$ with $T_{\mu \nu}$ the stress-energy tensor \cite{Garcia:2010xz}.

The modified gravity can also lead to a change in cosmic expansion rate in the early universe.  It, therefore, provides a solution to the horizon problem, flatness problem, and other problems related to observations of cosmic microwave background radiation \cite{Starobinsky:1980te}.  In this way, the modified gravity is one of mechanisms for changing the cosmic expansion rate in the very early and the present epochs.  It is, however, possible that effects of the modified gravity will be detected in cosmological observables which look to be consistent with astronomical observations at the moment.  If the cosmic expansion rate is different from that in the standard cosmological model during the big bang nucleosynthesis (BBN) epoch, primordial abundances of light elements can be different from those predicted in standard BBN (SBBN) model.  These differences may be detected in future observations of elemental abundances.  Therefore, it is meaningful to investigate effects of the modified gravity on BBN theoretically, and predict light element abundances in the modified gravity models.  We can also derive constraints on the modified gravity during the BBN epoch independently of those during the extremely early and present epoch.  Effects of the $f(R)$ gravity, where $f(R)$ is a general function of $R$, i.e., Ricci scalar, has been studied for a specific type of $f(R) \propto R^n$ as a simple extension of the general relativity \cite{Lambiase:2006dq,Kang:2008zi,Kusakabe:2015yaa}.  Primordial abundances of D, $^{3, 4}$He, and $^{6,7}$Li have been calculated in the model, and a constraint on the model has been derived from comparison of calculated abundances with observational data \cite{Kusakabe:2015yaa}.

In this paper, we analyze effects of the $f(G)$ modified gravity during BBN epoch, and derive constraints  on the model based on observed light element abundances.  In section \ref{sec2}, the $f(G)$ model is introduced.  We study a specific case of $\xi(G)=df(G)/dG \propto t^p$ with $t$ the cosmic time and $p$ a power-law index since numerical BBN calculations in the model are easily performed.  In section \ref{sec3}, our BBN code and treatment of modified gravity effects are explained.  In section \ref{sec4}, observational constraints on primordial light element abundances are described.  In section \ref{sec5}, we show results of BBN in the $f(G)$ gravity model, and derive constraints on the model.  In section \ref{sec6}, we summarize this study.  In this paper, we adopt natural units of $c=k_{\rm B} =1$, where $c$ is the light speed and $k_{\rm B}$ is the Boltzmann constant, and the convention of $[\nabla_\alpha, \nabla_\beta] A^\mu ={R^\mu}_{\nu \alpha \beta} A^\nu$ and $R_{\mu \nu} ={R^\lambda}_{\mu \lambda \nu}$.
 
\section{Model}\label{sec2}
\subsection{$f(G)$ gravity}\label{sec2_1}
Equations of motion in the $f(G)$ gravitational model are shown in this section.  The action taken in this paper is given by
\begin{equation}
\label{eq1}
S= \frac{1}{2 \kappa^2} \int d^4 x \sqrt{-g} \left[ R +\kappa^2 f(G) \right]+S_\mathrm{m} (g_{\mu \nu}, \phi_\mathrm{m} ),
\end{equation}
where
$\kappa^2 =8\pi {\mathcal G}$ is defined with ${\mathcal G}$ the Newton's constant,
$g_{\mu \nu}$ is the metric tensor,
$g$ is the determinant of the metric tensor,
$S_\mathrm{m}$ is the action of the matter field $\phi_\mathrm{m}$.
We assume the spatially flat Friedmann-Lema\^{i}tre-Robertson-Walker metric as the same as in the standard cosmological model
\begin{equation}
\label{eq2}
ds^2 = -dt^2 +a(t)^2 \left( dx^2 +dy^2 +dz^2 \right),
\end{equation}
where
$a(t)$ the scale factor of the universe.
For this metric, the nonzero components of the Ricci tensor are given by
\begin{eqnarray}
R_{00} &=& -3 \frac{\ddot{a}}{a}, \label{eq3}\\
R_{ij} &=& a^2 \left[ \frac{\ddot{a}}{a} + 2\left( \frac{\dot{a}}{a} \right)^2\right] \delta_{ij}. \label{eq4}
\end{eqnarray}
The Ricci scalar is given by
\begin{equation}
\label{eq5}
R = 6 \left[ \frac{\ddot{a}}{a} + \left( \frac{\dot{a}}{a} \right)^2 \right].
\end{equation}
The GB term is given by
\begin{eqnarray}
G &\equiv & R^2 -4 R_{\mu \nu} R^{\mu \nu} +R_{\mu \nu \rho \sigma} R^{\mu \nu \rho \sigma} \nonumber\\
&=& 24 H^2 \left( \dot{H} +H^2 \right). \label{eq6}
\end{eqnarray}

For a matter, on the other hand, we assume a perfect fluid described with an energy density $\rho(t)$ and pressure $p(t)$
\begin{equation}
\label{eq7}
{T^\mu}_\nu = \mathrm{diag}\left( -\rho, p, p, p \right).
\end{equation}
The field equation for the $f(G)$ gravity is then derived by varying the action (eq. \eqref{eq1}) with respect to the metric tensor,
\begin{eqnarray}
\label{eq8}
-\frac{g_{\mu \nu}}{2} R
+R_{\mu \nu} 
+\kappa^2 \left\{
-\frac{g_{\mu \nu}}{2} f(G)
+2\left(R_{\mu \nu} +D_{\mu \nu}\right) \left( f' R \right)
~~~~~~~~~~~~~~~~~~
\right.
&&
\nonumber \\
\left.
+8 \nabla^\lambda \left[ \nabla _\mu \left( f' R_{\nu \lambda} \right) + \nabla _\nu \left( f' R_{\mu \lambda} \right) \right]
~~~~~~~~~~~~~~~~~~
\right.
&&
\nonumber \\
\left.
-4 g_{\mu \nu} \nabla^\alpha \nabla^\beta \left(f' R_{\alpha \beta} \right)
-4 \Box \left( f' R_{\mu \nu} \right)
+2 f' R_{\mu \alpha} {R^{\alpha}}_\nu
\right.
&&
\nonumber \\
\left.
+2 f' {R_\mu}^{\alpha \beta \gamma} R_{\nu \alpha \beta \gamma}
+4 \nabla^\rho \nabla^\sigma \left( f' R_{\mu \rho \nu \sigma} +f' R_{\mu \sigma \nu \rho} \right)
\right\}
&=&
\kappa^2 T_{\mu \nu},
\end{eqnarray}
where $f'(G)=df(G)/dG$ is the derivative with respect to $G$, 
$\nabla_{\mu}$ is the covariant derivative operator,
$\Box \equiv g^{\mu \nu} \nabla_\mu \nabla_\nu =\nabla^\mu \nabla_\mu$ is the D'Alambertian operator, and the differential tensorial operator $D_{\mu \nu}$ is defined as
\begin{equation}
\label{eq9}
D_{\mu \nu} \equiv g_{\mu \nu} \Box - \nabla_\mu \nabla_\nu.
\end{equation}
Especially, varying the action with respect to the 0-0 and $i$-$i$ components of metric tensor, one derives
\begin{eqnarray}
\kappa^2 \left[ \frac{f(G) -G f'}{2} + 12 H^3 \dot{f'} \right] +3H^2 &=& \kappa^2 \rho \label{eq10} \\
\kappa^2 \left[ \frac{f(G) -G f'}{2} + 4 H^2 \left( \ddot{f'} + 2 H \dot{f'} \right) +8H \dot{H} \dot{f'} \right] + 2 \dot{H} +3 H^2 &=& - \kappa^2 p, \label{eq11}
\end{eqnarray}
where
$H\equiv \dot{a}/a$ is the expansion rate of the universe.

In addition, the conservation of the stress-energy tensor $\nabla^\mu T_{\mu \nu}$ holds, which results in the equation
\begin{equation}
\label{eq12}
\dot{\rho} +3 H \left( \rho +p \right) =0.
\end{equation}
This equation can be found using eqs. \eqref{eq10} and \eqref{eq11} also.

\subsection{Assumption on $f(G)$}\label{sec2_2}

In this paper, we constrain the model space of $f(G)$.   We define a variable $\xi \equiv f'$, and derive equations from eqs. \eqref{eq10} and \eqref{eq11},
\begin{eqnarray}
g(H; \xi, \rho) &=& 2 \kappa^2 \rho -6 H^2 -24 H^3 \kappa^2 \dxi - \kappa^2 f(G) +\kappa^2 \xi G=0 \label{eq13}\\
\dot{H}(H; \xi, \rho, p) &=& \frac{ \kappa^2 \left[ 4H^2 \left( H \dot{\xi} - \ddot{\xi} \right) - \left( \rho +p \right)\right]}{2 \left( 1+ 4 \kappa^2 H \dot{\xi}\right)}. \label{eq14}
\end{eqnarray}

We assume that the $\xi$ term scales as a power-law function of time, i.e.,
\begin{equation}
\label{eq15}
\xi(t) = \xi_0 \left( t /t_0 \right)^p,
\end{equation}
where
$\xi(t)$ is dimensionless, 
$\xi_0 =\xi(t_0)$ is dimensionless parameter for the $\xi$ value at $t_0 =1$ s, and
$p$ is a power-law index.
This model is a toy model for time-dependent $f(G)$ term.  In this paper, we show how to solve BBN in the $f(G)$ modified gravity model exactly in numerical calculation.  Usually, detailed BBN calculations are performed with time taken as the evolution parameter.  Because of this assumption of the explicit time dependence of $\xi$, values of $\xi$, $\dot{\xi}$, and $\ddot{\xi}$ can be specified before solving $H(t)$, $\dot{H}(t)$, $G(t)$, and $f(G)$ using eqs. \eqref{eq6}, \eqref{eq13}, \eqref{eq14}, and \eqref{eq17}.  Since we do not need to treat the parameter $\xi$ as unknown in solving the system of equations, it becomes somewhat simple to solve physical variables in the present model.  It would be possible to solve BBN in other models of $f(G)$ gravity based on the method described below.  However, methods of the solution can be more complicated depending on the adopted models, and it is beyond the scope of this study to investigate many possible models of $f(G)$ function.

Effects of the $\xi$ term on primordial light element abundances are constrained from observations of the primordial abundances.  Since the primordial abundances are sensitive to the cosmic expansion history only in the BBN epoch, the GB term during BBN can be constrained.  We then assume that the GB term exists in the temperature range of $100 \geq T_9 =T/(10^9~{\rm K}) \geq 0.01$.

\section{BBN calculation with modified expansion rate}\label{sec3}

The public BBN calculation code \cite{Kawano1992,Smith:1992yy} is utilized and modified.  In this study, the effective number of neutrino species is assumed to be three.  We updated reaction rates of nuclei with mass numbers $\le  10$ using the JINA REACLIB Database \cite{Cyburt2010} (the latest version taken in December, 2014).  The neutron lifetime is the central value of the Particle Data Group, $880.3 \pm 1.1$~s~\cite{Agashe:2014kda}.  The baryon-to-photon ratio is taken from the accurate value, $(6.037 \pm 0.077) \times 10^{-10}$ \cite{Ishida:2014wqa}, corresponding to the baryon density in the base $\Lambda$CDM model (Planck+WP+highL+BAO) determined from Planck observation of cosmic microwave background, $\Omega_\mathrm{m} h^2 =0.02205 \pm 0.00028$ \cite{Ade:2013zuv}.

In general, two independent equations among equations of motion are used in BBN numerical codes.  For example, in the Kawano's code, equations of the Hubble expansion rate $H$ and the time derivative of temperature $dT/dt$ are used.  In the present modified gravity model, the Hubble expansion rate is given by eq. \eqref{eq13}.  The energy conservation equation, i.e., eq. \eqref{eq12}, is, however, the same as that in the SBBN model.  We can, therefore, use the same equation for the time evolution of temperature as that in the SBBN model (eq. (D.26) in ref. \cite{Kawano1992}).  Then, only one modification of the Hubble rate should be added to the code for BBN network calculations.  We should solve the Hubble expansion rate $H(t)$ and $\dot{H(t)}$, and calculate the function
\begin{equation}
\label{eq16}
f(G) =\int_{G_i}^{G} \xi(t) dG.
\end{equation}

We assume that the $\xi$ term exists in the temperature range of $T_9 =[10^2, 10^{-2}]$.  This temperature range corresponds to the time range of $t ={\mathcal O} (10^{-2} -10^6)$ s in SBBN model although the time range depends on the model parameters in the $f(G)$ model.  The most important temperature relevant to SBBN is $T_9 \sim 1$.  The power-law of $\xi(t) =\xi_0 (t/t_0)^p$ is assumed with the range of index $p=[-2, 6]$.  This range is chosen since it includes critical values of $p=2$ and $4$, and effects of $f(G)$ gravity outside the range are trivial (see section \ref{sec5}).  Both of the positive and negative $\xi$ cases are considered.  Another parameter is the initial value of the $f(G)$ term, i.e., $f(G_i)$.  The initial time corresponding to the initial temperature of $T_9 =10^2$ is assumed to be the same as that of SBBN.  This choice of initial time only affects the normalization of the $\xi$ value and is therefore not important.

We adopt the Newton-Raphson method to search for a correct solution with careful attention to the possibility of finding fake solutions.  We used the following technique.  For a given time, the temperature is calculated, and various physical quantities including the the energy density and pressure are derived.  Since the function $\xi(t)$ has been given (eq. \eqref{eq15}), the $\dot{H}$ value is given as a function of $H$ (eq. \eqref{eq14}).  The GB term $G$ is then given as a function of $H$ also (eq. \eqref{eq6}).  The function $f(G)$ is calculated by integration of $\xi(t)$ with respect to the GB term (eq. \eqref{eq16}). The value of $G$ at time $t+\Delta t$ is estimated with the equation,
\begin{equation}
\label{eq17}
f(G(t+\Delta t)) =f(G(t)) + \frac{\xi(t) +\xi(t +\Delta t)}{2} \left[ G(t+\Delta t) -G(t) \right],
\end{equation}
where
$f(G(t))$ and $\xi(t)$ is the values of $f(G)$ and $\xi$ at the previous time in the calculation.  The function $g(H; \xi, \rho)$ is then given as a function of $H$, and a solution of $g(H; \xi, \rho)=0$ is searched for.  In the numerical calculation, the following equations are utilized.
\begin{eqnarray}
\frac{dg}{dH} &=& -12 H +\kappa^2 \left[-72 H^2 \dxi + \xi \frac{dG}{dH}- \frac{df(G)}{dH} \right] \label{eq18} \\
\frac{dG}{dH} &=& 96 H^3 +48 H \dot{H} +24 H^2 \frac{d\dot{H}}{dH} \label{eq19} \\
\frac{d \dot{H}}{dH} &=& \frac{2 \kappa^2 \left( 3H^2 \dxi -2 H \ddxi \right)}{1 +4 \kappa^2 H \dxi} -\frac{4 \kappa^2 \dxi}{1+4 \kappa^2 H \dxi} \dot{H} \label{eq20} \\
\frac{d f(G)}{dH} &=& \frac{\xi(t) +\xi(t +\Delta t)}{2} \frac{dG}{dH}. \label{eq21}
\end{eqnarray}

At the initial time in the calculation, the value for an initial guess of Hubble rate is given by $H_{\rm guess}=H_{\rm SBBN}$, where $H_{\rm SBBN}$ is the rate in SBBN model at the same temperature.  At later times, the initial guesses are given by solutions found at the previous times in the calculation.  The equation of the Hubble rate (eq. \eqref{eq13} with eqs. \eqref{eq6}, \eqref{eq14}, and \eqref{eq17}) is a transformed quintic equation.  Their coefficients, however, evolve as a function of time.  Therefore, the number of real roots can change with time.  It is then possible that a real root for the Hubble rate evolves to a complex root at some time.  This root is of course inappropriate solution for the expansion rate of the real universe.  Also if a root becomes negative at some time, it is inappropriate.  Therefore, we need to evaluate whether there is a real positive solution for a parameter set of ($\xi_0, p, f(G_i)$).  We classify situations of the root finding into six cases:  (1) a positive real root is found and it is not different from the previous root by more than three orders of magnitude, (2) any real root is not found, (3) a negative real root is found, (4) a positive real root is found but is is larger than that in SBBN by a factor of more than $10^8$, (5) a positive real root is found but it is different from the previous root by more than three orders of magnitude, (6) a positive real root is found which is different from case (5), but the sign of $df/dH$ is opposite to that at the previous time.

We regard that an appropriate root is obtained only in the case (1).  In cases (2)--(6), the solution most probably becomes negative, or complex, or too large so that they are included in a ``no solution'' region.  The case (2) can be caused by the disappearance of the real solution.  In this case, we do not have a smooth positive solution which continuously exists during the temperature range for the calculation.  The case (3) can be caused by a decrease of the root value below zero.  The case (4) can be caused by an unrealistically large root, and was excluded from real solutions.  The case (5) can be caused by finding a root different from that in the previous time, or a change of the root value across zero.  The case (6) can be caused by finding a different root.  It indicates that one positive real root becomes negative, and the code find another positive real root next to it, most probably.

\section{Observed light element abundances}\label{sec4}
Calculated BBN results are compared to the following observational constraints on light element abundances.

The primordial $^4$He abundance is estimated with observations of metal-poor extragalactic
H II regions.  We use the latest determination of $Y_{\rm p}=0.2551\pm 0.0022$~\cite{Izotov:2014fga}.  When the central values of adopted reaction rates, the neutron lifetime, and the baryon-to-photon ratio are used, the calculated abundances in the SBBN model is out of the $2\sigma$ observational limit.  The $4\sigma$ range is then adopted in this study.

The primordial D abundance is estimated with observations of metal-poor Lyman-$\alpha$ absorption systems in the foreground of quasi-stellar objects.  We use the weighted mean value of D/H$=(2.53 \pm 0.04) \times 10^{-5}$~\cite{Cooke:2013cba}, and adopt its $4\sigma$ range since the $2\sigma$ range is inconsistent with the theoretical abundances in the SBBN model, similarly to the $^4$He abundance.

$^3$He abundances are measured in Galactic H II regions through the $8.665$~GHz
hyperfine transition of $^3$He$^+$ ion.  These are not the primordial abundance but present values which have contributions from Galactic chemical evolution taking into account production and destruction of nuclei in stars.  Nevertheless, it is very hard to reduce elemental abundances significantly in standard Galactic chemical evolution theory.  We then adopt the $2\sigma$ upper limit from the abundance $^3$He/H=$(1.9\pm 0.6)\times 10^{-5}$~\cite{Bania:2002yj} in Galactic H II regions, as a rough guide. 

The primordial $^7$Li abundance is estimated with observations of Galactic metal-poor stars.  We use the abundance $\log(^7$Li/H)$=-12+(2.199\pm 0.086)$ derived in a 3D nonlocal thermal equilibrium model~\cite{Sbordone2010}, and adopt its $2\sigma$ range.

$^6$Li abundances in Galactic metal-poor stars have also been measured.   We adopted the least stringent $2~\sigma$ upper limit of all limits for stars reported in \cite{Lind:2013iza}, i.e., $^6$Li/H=$(0.9\pm 4.3)\times 10^{-12}$ for the G64-12 (nonlocal thermal equilibrium model with 5 free parameters).

\section{Result}\label{sec5}

\subsection{Case of $\xi_0 >0$ and $f(G_\mathrm{i})=0$}\label{sec5_1}

\begin{figure}[tbp]
\begin{center}
\includegraphics[width=8.0cm,clip]{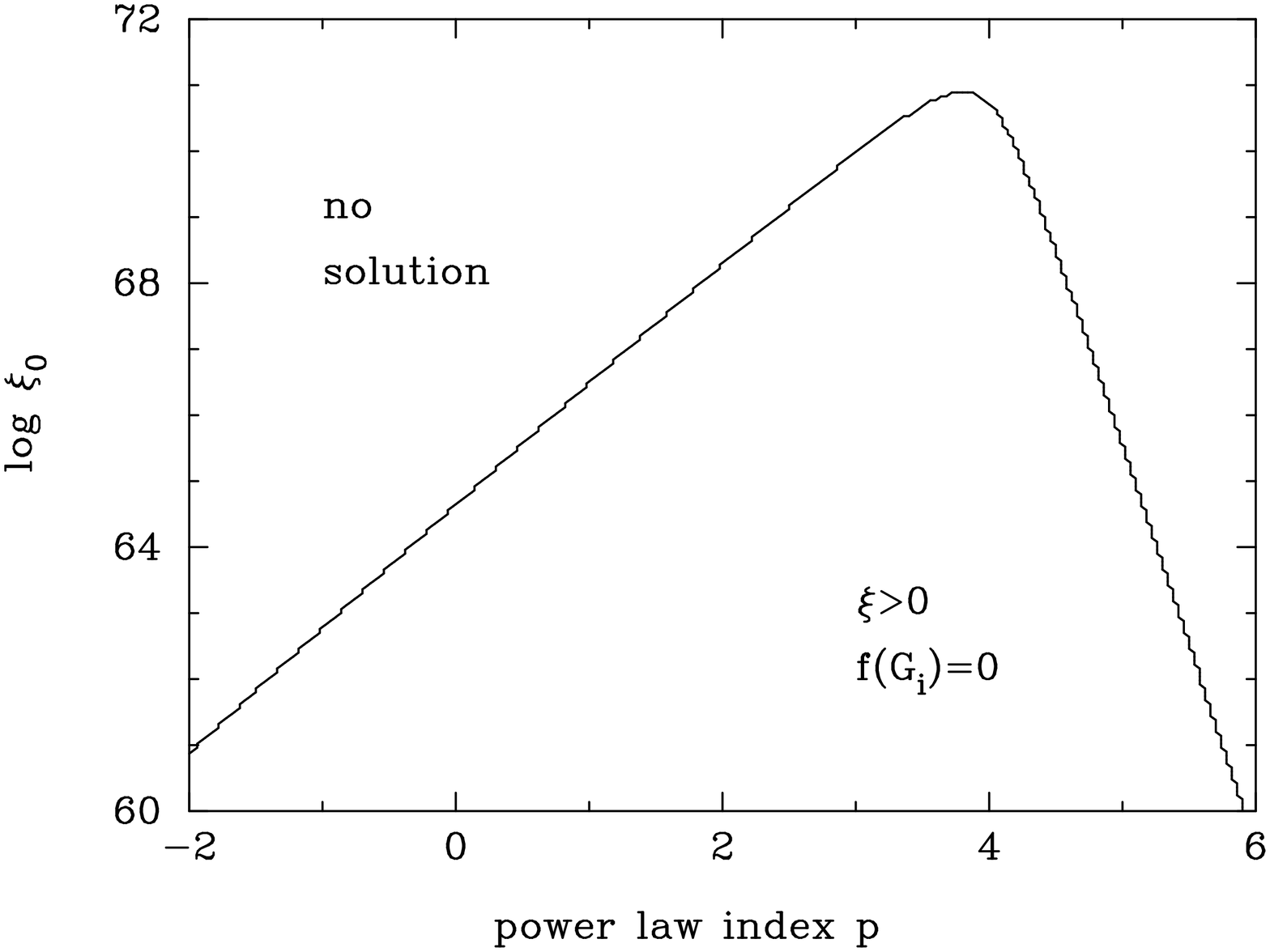}
\caption{Boundary of regions with and without proper cosmological solutions in the parameter plane of ($p$, $\xi_0$) in the case of $\xi_0>0$.  For this figure the initial value of $f(G)$ is fixed to be $f(G_\mathrm{i})=0$.  The region marked with `no solution' above the boundary corresponds to parameter sets which have no proper solution of cosmic expansion.  This region is therefore excluded. \label{fig1}}
\end{center}
\end{figure}

Figure \ref{fig1} shows boundary of regions with and without proper cosmological solutions in the parameter plane of ($p$, $\xi_0$) in the case of $\xi_0>0$.  The initial value of $f(G)$ is fixed as $f(G_\mathrm{i})=0$ for purposes of illustration.  The region marked with `no solution' above the boundary corresponds to parameter sets which lead to no proper solution of cosmic expansion in the BBN calculation.  This region is therefore excluded.  We find that the constraint in the parameter plane of ($p$, $\xi_0$) for $\xi_0>0$ is predominantly given by the continuous existence of the solution $H>0$ during the BBN epoch.  Final abundances of light elements are within the adopted observational limits, in the parameter region below the boundary.  The limits on light element abundances are, therefore, less constraining than the existence of a solution.


The reason for the shape of the boundary is explained as follows.  We consider deviations of the cosmic expansion rate from that in the standard cosmological model.  First, some characteristics of the standard model are reviewed.   The expansion rate during the radiation dominated epoch is given by $H=\dot{a}/a=1/(2t)$.  Also the following equations holds,
\begin{eqnarray}
H^2 &= & \frac{\kappa^2}{3} \rho \label{eq22} \\
\dot{H} &=& -\frac{\kappa^2}{2} \left( \rho +p \right) \label{eq23} \\
\dot{H} +H^2 &=& -\frac{\kappa^2}{6} \left( \rho +3p \right) \label{eq24}
\end{eqnarray}
Since $\dot{H}+H^2 <0$ is always satisfied, the inequality $G<0$ holds (eq. \eqref{eq6}).  The time derivative of $G$ is given by
\begin{equation}
\label{eq25}
\frac{dG}{dt} =-\frac{4}{3} \kappa^4 \frac{d}{dt} \left[ \rho \left( \rho +3p \right)\right].
\end{equation}
Therefore, the inequality $dG/dt >0$ is satisfied.  

If the deviation of expansion rate is not large, it follows that
\begin{eqnarray}
-6H^2 &^\propto_\sim & -t^{-2} \label{eq26} \\
-24 H^3 \kappa^2 \dxi &^\propto_\sim & - H^3 \dxi \propto - t^{-3} \left(p \xi_0 t^{p-1}\right) =-p \xi_0 t^{p-4} \label{eq27} \\
-\kappa^2 f(G) &^\propto_\sim & - \int \xi dG \propto - \int \left( \xi_0 t^p \right) d(-t^{-4}) \propto - \xi_0 \int \xi_0 t^{p-5}dt \propto - \xi_0 \frac{\left[ t^{p-4} \right]^t_{t_\mathrm{i}}}{p-4} \label{eq28} \\
\kappa^2 \xi G &^\propto_\sim & \xi G \propto \left( \xi_0 t^p \right) (-t^{-4}) = -\xi_0 t^{p-4}. \label{eq29}
\end{eqnarray}
We note that once the expansion rate deviate from the value in the standard model significantly, the relations of $H \propto t^{-1}$ and $G \propto -t^{-4}$ are broken and the above scalings no longer hold.

\paragraph{$p \lesssim 4$:}
As $t$ increases, amplitudes of the second, third, and fifth terms after the first equality in eq. \eqref{eq13} decrease.  However, the amplitude of the fourth term does not change much since the early time of $t\sim t_\mathrm{i}$ contributes to the integral (eq. \eqref{eq28}) predominantly.  As a result, the fourth term becomes more and more important relatively.  Because of $f(G)>0$ for $\xi_0 >0$ (eq. \eqref{eq28}), the fourth term is negative and decelerates the cosmic expansion.  The $f(G)$ term, therefore, works similarly to a negative dark energy $\Lambda$.  When this term becomes large, we lose a real positive root of the expansion rate which smoothly connects to the root in the standard model.  In the parameter region marked as `no solution', the solution is lost until the cosmic temperature decreases down to the final value of $T_9 =10^{-2}$.  On the boundary of the `no solution' region, solutions disappear right before the temperature decreases to $T_9 =10^{-2}$.  On the boundary, therefore, the equation is satisfied,
\begin{equation}
\label{eq30}
f(G) \approx 2\rho,
\end{equation}
at $T_9 =10^{-2}$.  This equation holds since the second, third, and fifth terms in eq. \eqref{eq13} are negligible for $H\rightarrow 0$.
When the power-law index $p$ is larger, the amplitude of the fourth term $\propto |\left[ t^{p-4} \right]^t_{t_\mathrm{i}}| \sim t_\mathrm{i}^{p-4}$ is smaller.  Therefore, the effect of the $f(G)$ term is smaller and larger amplitudes of $\xi_0$ are allowed from observational constraints.  This is the reason for the upward-sloping curve for $p \lesssim 4$.

\paragraph{$p \gtrsim 4$:}

As $t$ increases, amplitudes of the third to fifth terms in eq. \eqref{eq13} increase, while that of the second term decreases.  Since all of the third to fifth terms are negative, their sum works similarly to a negative $\Lambda$.  The boundary then corresponds to the point of eq. \eqref{eq30} also for $p \gtrsim 4$.  All of these terms roughly scale as $t^{p-4}$ for $t \gg t_\mathrm{i}$ (eqs. \eqref{eq27}--\eqref{eq29}).  When the power-law index $p$ is larger, the quantity $t^{p-4}$ is larger so that the effects of the three terms are larger.  As a result, smaller amplitudes of $\xi_0$ are allowed.  This results in the downward-sloping curve for $p \gtrsim 4$.

We note that a magnitude of the parameter $\xi_0$ allowed from the BBN constraint is typically huge.  For example, we consider a case of $p=4$ in which the cosmic expansion rate is not so much different from that in the standard cosmological model.  The third to fifth terms in eq. \eqref{eq13} then become constants (see eqs. \eqref{eq27}--\eqref{eq29}).  Since the three terms scale similarly, we just take the fifth term for simplicity.  Then, the amplitude of $\xi$ which significantly affects the expansion rate is estimated as follows.  In this case, the first, the second and the fifth terms in eq. \eqref{eq13} are of the same order of magnitude, i.e.,
\begin{equation}
\label{eq_add1}
2 \kappa^2 \rho \sim 6H^2 \sim \kappa^2 |\xi||G|.
\end{equation}
The GB term in the standard model is given by
\begin{equation}
\label{eq_add2}
G =24 H^2 \left(\dot{H} +H^2\right) \sim-\frac{8}{3} \kappa^4 \rho^2,
\end{equation}
where eqs. \eqref{eq6}, \eqref{eq22}, and \eqref{eq24} were used, and
the radiation dominated epoch is assumed.  The energy density in the radiation dominated universe is given by
\begin{equation}
\label{eq_add3}
\rho =\frac{\pi^2}{30} g_\ast T^4,
\end{equation}
where
$g_\ast$ is the relativistic degrees of freedom for the energy density \cite{kolb1990}.
Inserting eqs. \eqref{eq_add2} and \eqref{eq_add3} into $|\xi|\sim 2\rho/|G|$ from eq. \eqref{eq_add1}, we obtain
\begin{equation}
\label{eq_add4}
|\xi| \sim \frac{3}{4} \frac{1}{\kappa^4 \rho} =\frac{45}{128 \pi^4 g_\ast} \left( \frac{m_{\rm Pl}}{T}\right)^4 ={\mathcal O}(10^{84}) \left(\frac{10.75}{g_\ast}\right) \left(\frac{1~{\rm MeV}}{T}\right)^4,
\end{equation}
where
$m_{\rm Pl}={\mathcal G}^{-1/2}=\sqrt{8\pi}/\kappa=1.22\times 10^{19}$ GeV is the Planck mass.  Since the factor $(m_{\rm Pl}/T)^4$ is huge during the BBN epoch of $T\sim 1$ MeV, constraints on the present modified gravity model exclude huge amplitudes of $\xi$.

\subsection{Case of $\xi_0 <0$ and $f(G_\mathrm{i})=0$}\label{sec5_2}

\subsubsection{Parameter search}\label{sec5_2a}

\begin{figure}[tbp]
\begin{center}
\includegraphics[width=8.0cm,clip]{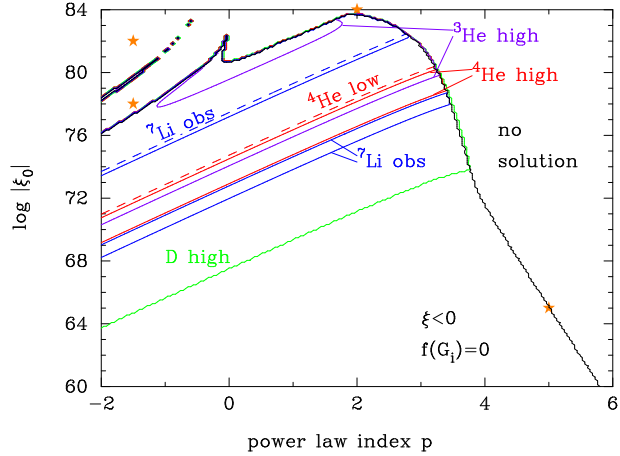}
\caption{Same boundary (black line) as in figure \ref{fig1}, but in the parameter plane of ($p$, $|\xi_0|$) for $\xi_0<0$.  Contours of calculated light element abundances are also shown.  Solid and dashed lines for D (green line), $^3$He (purple lines), and $^4$He (red lines) correspond to the observational upper (`high') and lower (`low') limits, respectively on their abundances.  Blue solid and dashed lines marked with `$^7$Li obs' correspond to observational upper and lower limits, respectively, on $^7$Li/H abundances.  Four orange stars located in the `no solution' region correspond to the parameter set analyzed in section \ref{sec5_2b}.\label{fig2}}
\end{center}
\end{figure}

Figure \ref{fig2} shows the same boundary (black line) as in figure \ref{fig1} but in the parameter plane of ($p$, $|\xi_0|$) for $\xi_0<0$.  Contours of calculated light element abundances are also drawn.  Solid and dashed lines for D (green line), $^3$He (purple lines), and $^4$He (red lines) correspond to the observational upper (`high') and lower (`low') limits, respectively, on their abundances.  Blue solid and dashed lines marked with '$^7$Li `obs' correspond to observational upper and lower limits, respectively, on $^7$Li/H abundances.   We find that observational limits on primordial abundances give important constraints in the region of $p \lesssim 4$, contrary to the case of $\xi_0>0$.


The shapes of the solution boundary and the abundance contours are explained as follows.  Deviations of the cosmic expansion rate from that in the standard cosmology is considered again.  

\paragraph{$p \lesssim 2$:}

As $t$ increases, the fourth term in eq. \eqref{eq13} gradually becomes dominant.  Because of $f(G)<0$ for $\xi_0 <0$, the fourth term is positive (eq. \eqref{eq28}).  Therefore, the term causes an acceleration of the cosmic expansion similarly to a positive $\Lambda$.  When this term becomes large, the expansion rate deviates from the standard rate and the universe inflates.  Since the fourth term scales as $\sim -t_\mathrm{i}^{p-4}$, a larger power-law index $p$ gives smaller amplitudes of the fourth term.  The effect of the $f(G)$ term is, therefore, smaller and larger $|\xi_0|$ are allowed.  This fact explains the upward-sloping curves for the solution boundary and the abundance constraints for $p \lesssim 2$.  For a fixed power-law index $p$, larger $|\xi_0|$ values causes larger $H$ values or faster cosmic expansion.  Effects of the $f(G)$ gravity on light element abundances are, therefore, larger.  Reasons of the complicated shape of the boundary are described below in section \ref{sec5_2c}. 

\paragraph{$2 \lesssim p \lesssim 4$:}

As $t$ increases, amplitudes of components which are proportional to $t^{p-4}$ in the third to fifth terms in eq. \eqref{eq13} become larger relative to the term $-6H^2 \propto t^{-2}$.  In a late time of BBN, then a large deviation of the expansion rate is induced, and a proper solution possibly disappears (see section \ref{sec5_2b}).  A larger power-law index $p$ gives larger amplitudes of the third to fifth terms. Since the effect of modified gravity is thus larger, smaller $|\xi_0|$ values are allowed.  For this reason, the solution boundary is downward-sloping in the region $2 \lesssim p \lesssim 4$.  

\paragraph{$p \gtrsim 4$:}

As $t$ increases, amplitudes of the third to fifth terms in eq. \eqref{eq13} become large relative to that of the second term.  The third to fifth terms are positive so that their sum works similarly to a positive $\Lambda$.  When the amplitude $|\xi_0|$ is larger than a critical value, a real positive root of $H$ is lost (see section \ref{sec5_2b}).  All of these terms roughly scale as $t^{p-4}$ for $t \gg t_\mathrm{i}$.  When the power-law index $p$ is larger, the quantity $t^{p-4}$ is larger and the effects of the three terms are larger.  Smaller $|\xi_0|$ values are then allowed.  This is the reason for the downward-sloping curve for $p \gtrsim 4$, similarly to the case of $\xi_0>0$.

\subsubsection{Examples of cosmic evolution}\label{sec5_2b}

Evolutions of the expansion rate are illustrated for four parameter sets in the `no solution' region, which are indicated by orange stars in figure \ref{fig2}.  Since the boundary of the existence of proper solution has a complicated shape, we check the cosmic histories for four typical parameter cases of `no solution' region.  As seen below, the cosmic expansion rates evolve much differently depending on the parameter sets.

\paragraph{($p$, $\log |\xi_0|$) =($5$, $65$):}
As the time increases, the amplitudes of the third to fifth terms in eq. \eqref{eq13} rapidly increase with respect to that of the second term.  The local minimal value in the function $g(H; \xi, \rho)$ (eq. \eqref{eq13}) as a function of $H$ then increases.  At a certain time, the local minimal value becomes positive, and a real positive root for $H$ disappears.  We note that the amplitudes of the third to fifth terms are not larger than that of the second when the solution is lost although the ratios of amplitudes of former three terms and the latter term are rapidly increasing.

\begin{figure}[tbp]
\begin{center}
\includegraphics[width=8.0cm,clip]{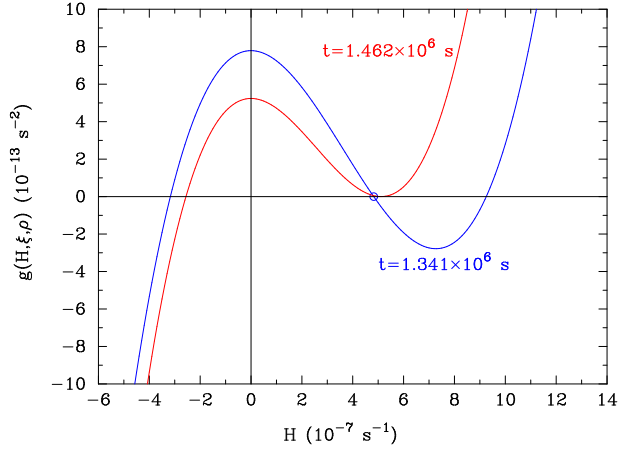}
\caption{The functions $g(H, \xi, \rho)$ as a function of $H$ for the parameter set of ($p$, $\log |\xi_0|$, $f(G_\mathrm{i})$)=($5$, $65$, $0$).  One line is the function at the cosmic time of $t=1.341 \times 10^6$ s which has a real positive solution marked by an open circle, while another is the function at $t =1.462 \times 10^6$ s after the real solution disappeared.  \label{fig_no_1a}}
\end{center}
\end{figure}

Figure \ref{fig_no_1a} shows the functions $g(H, \xi, \rho)$ as a function of $H$ for the parameter set of ($p$, $\log |\xi_0|$, $f(G_\mathrm{i})$)=($5$, $65$, $0$).  Two lines for $t=1.341 \times 10^6$ and $t =1.462 \times 10^6$ s are shown.  The former time has a real positive solution marked by an open circle, while the latter corresponds to the time right after the real solution disappeared.


\begin{figure}[tbp]
\begin{center}
\includegraphics[width=8.0cm,clip]{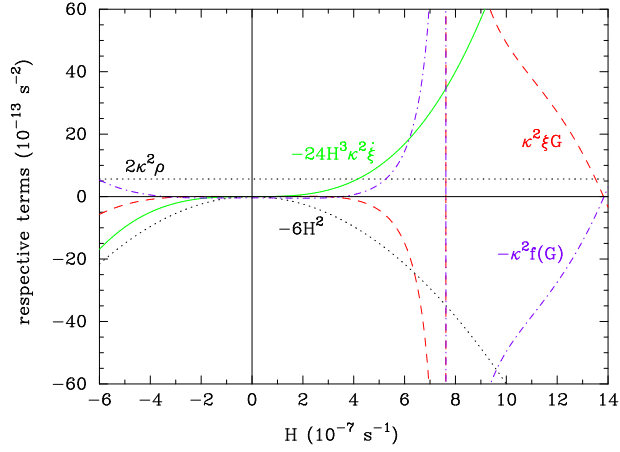}
\caption{Respective terms in the functions $g(H, \xi, \rho)$ (eq. \eqref{eq13}) at $t =1.462 \times 10^6$ s  as a function of $H$ for the same parameter set as in figure \ref{fig_no_1a}.  Dotted lines show the first and the second terms, while the solid, dot-dashed, and dashed lines show the third, fourth, and fifth terms, respectively.  \label{fig_no_1b}}
\end{center}
\end{figure}

Figure \ref{fig_no_1b} shows respective terms in the function $g(H, \xi, \rho)$ (eq. \eqref{eq13}) at $t =1.462 \times 10^6$ s as a function of $H$ for the same parameter set as in figure \ref{fig_no_1a}.  Dotted lines show the first and the second terms which exist in the standard cosmological model, while the solid (third term), dot-dashed (fourth), and dashed (fifth) lines show terms which exist only in the modified gravity model.  This figure is for a time right after the proper solution is lost (figure \ref{fig_no_1a}).  The vertical lines at $H/(10^{-7}~{\rm s}^{-1}) \sim 7.6$ correspond to asymptotic lines of $-\kappa^2 f(G)$ (fourth term) and $\kappa^2 \xi G$ (fifth).  The loss of a solution is thus caused in the `no solution' region by the three additional terms of the modified gravity in the function $g(H, \xi, \rho)$.  As seen in this case, a strong cancellation of the three terms can occur.


\paragraph{($p$, $\log |\xi_0|$) =($2$, $84$):}
In the epoch when the solution of $H$ is lost in the BBN calculation, the second and fifth terms in eq. \eqref{eq13} are negative while the third and fourth terms are positive.  The amplitudes of the modified-gravity terms satisfies $|24 H^3 \dxi| > |f(G)| > |\xi G|$.  We note that relations in the standard model (eqs. \eqref{eq22}--\eqref{eq24}) are not satisfied in this case since the deviation in the expansion rate from standard case is large.  Especially, the inequality $G>0$ or $-\dot{H}<H^2$ (eq. \eqref{eq6}) is satisfied in this epoch.  The amplitude of the first term is negligible compared to other terms.  At a certain time, the local minimal value in the function $g(H; \xi, \rho)$ becomes larger than zero, and a real positive root of $H$ disappears.

\begin{figure}[tbp]
\begin{center}
\includegraphics[width=8.0cm,clip]{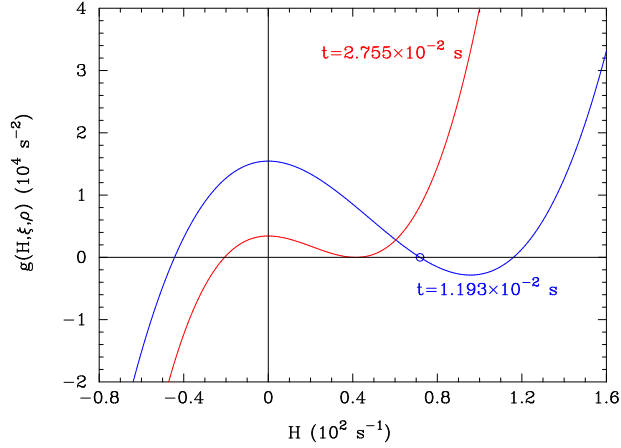}
\caption{The functions $g(H, \xi, \rho)$ as a function of $H$ for the parameter set of ($p$, $\log |\xi_0|$, $f(G_\mathrm{i})$)=($2$, $84$, $0$).  Two lines correspond to the functions at $t=1.193 \times 10^{-2}$ s with a real positive solution marked by an open circle and at $t =2.755 \times 10^{-2}$ s after the real solution disappeared.  \label{fig_no_3a}}
\end{center}
\end{figure}

Figure \ref{fig_no_3a} shows the functions $g(H, \xi, \rho)$ as a function of $H$ for the parameter set of ($p$, $\log |\xi_0|$, $f(G_\mathrm{i})$)=($2$, $84$, $0$).  The functions are plotted for $t=1.193 \times 10^{-2}$ s with a real positive solution marked by an open circle, and for $t =2.755 \times 10^{-2}$ s after the real solution disappeared.


\begin{figure}[tbp]
\begin{center}
\includegraphics[width=8.0cm,clip]{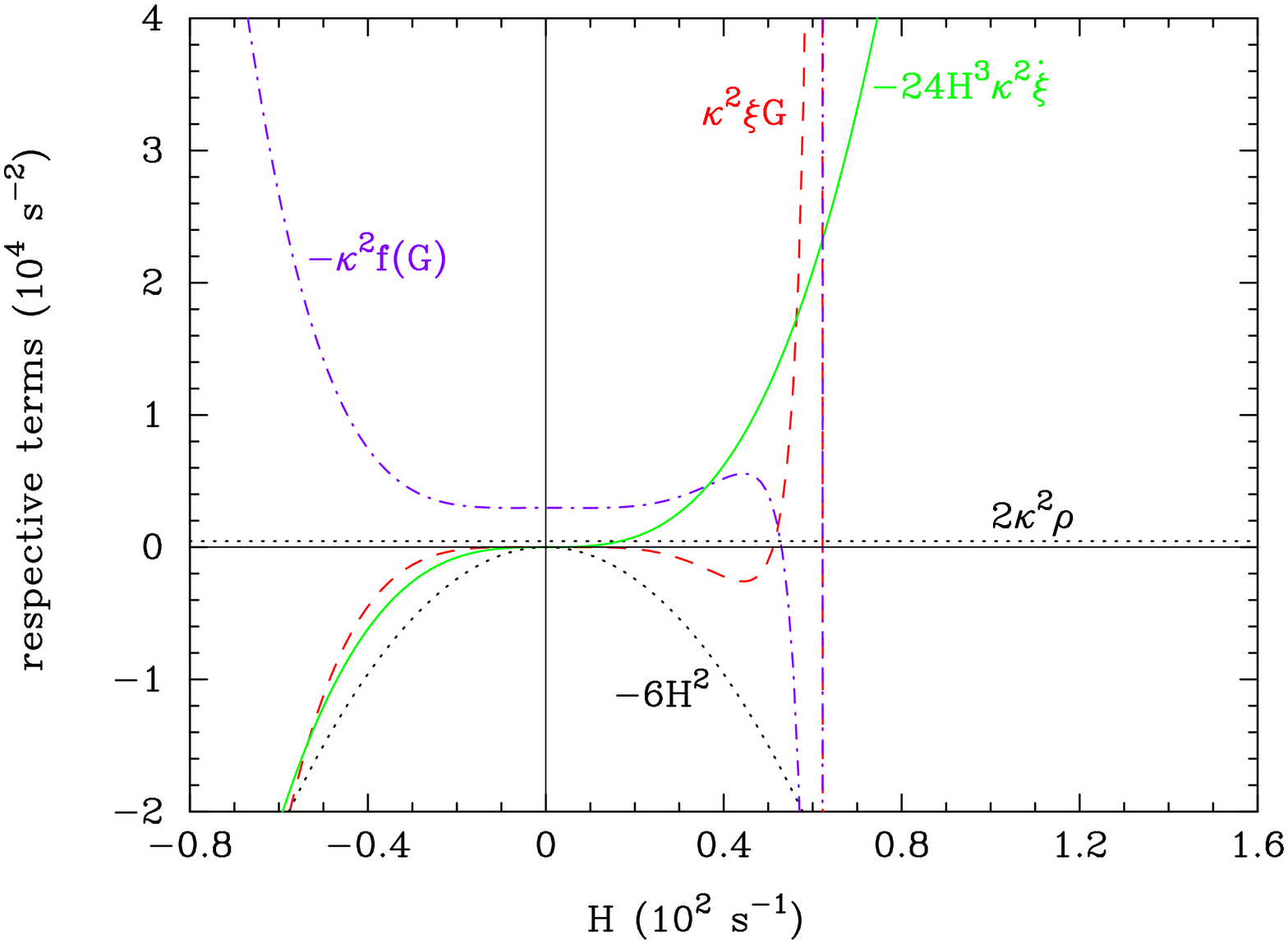}
\caption{Respective terms in the functions $g(H, \xi, \rho)$ (eq. \eqref{eq13}) at $t =2.755 \times 10^{-2}$ s as a function of $H$ for the same parameter set as in figure \ref{fig_no_3a}.  The lines correspond to the same quantities as in figure \ref{fig_no_1b}.  \label{fig_no_3b}}
\end{center}
\end{figure}

Figure \ref{fig_no_3b} shows respective terms in the function $g(H, \xi, \rho)$ (eq. \eqref{eq13}) at $t =2.755 \times 10^{-2}$ s as a function of $H$ for the same parameter set as in figure \ref{fig_no_3a}.  The lines correspond to the same quantities as in figure \ref{fig_no_1b}.  The vertical lines at $H/(10^{2}~{\rm s}^{-1}) \sim 0.6$ correspond to asymptotic lines of $-\kappa^2 f(G)$ (fourth term) and $\kappa^2 \xi G$ (fifth).


\paragraph{($p$, $\log |\xi_0|$) =($-1.5$, $78$):}
When the solution of $H$ is lost, the first and fourth terms are dominant, and are balanced with each other, and the second term, $-6H^2$, is much smaller than the two terms. The term $\kappa^2 f(G)$, however, becomes larger than the term $2\kappa^2 \rho$ with increasing time.  The local maximal value of the $g(H; \xi, \rho)$ function then becomes negative, and the proper root of $H$ disappears.

\begin{figure}[tbp]
\begin{center}
\includegraphics[width=8.0cm,clip]{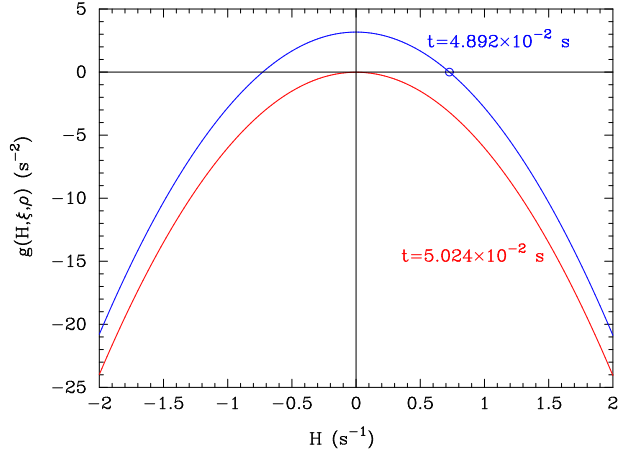}
\caption{The functions $g(H, \xi, \rho)$ as a function of $H$ for the parameter set of ($p$, $\log |\xi_0|$, $f(G_\mathrm{i})$)=($-1.5$, $78$, $0$).  Two lines correspond to the functions at $t=4.892 \times 10^{-2}$ s with a real positive solution marked by an open circle and at $t =5.024 \times 10^{-2}$ s after the real solution disappeared.  \label{fig_no_6a}}
\end{center}
\end{figure}

Figure \ref{fig_no_6a} shows the functions $g(H, \xi, \rho)$ as a function of $H$ for the parameter set of ($p$, $\log |\xi_0|$, $f(G_\mathrm{i})$)=($-1.5$, $78$, $0$).  We plot the functions at $t=4.892 \times 10^{-2}$ s with a real positive solution marked by an open circle, and at $t =5.024 \times 10^{-2}$ s after the real solution disappeared.


\begin{figure}[tbp]
\begin{center}
\includegraphics[width=8.0cm,clip]{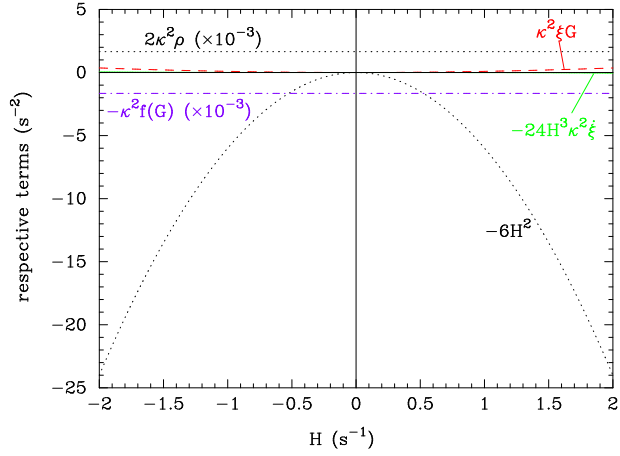}
\caption{Respective terms in the functions $g(H, \xi, \rho)$ (eq. \eqref{eq13}) at $t =5.024 \times 10^{-2}$ s as a function of $H$ for the same parameter set as in figure \ref{fig_no_6a}.  The lines correspond to the same quantities as in figure \ref{fig_no_1b}.  \label{fig_no_6b}}
\end{center}
\end{figure}

Figure \ref{fig_no_6b} shows respective terms in the function $g(H, \xi, \rho)$ (eq. \eqref{eq13}) at $t =5.024 \times 10^{-2}$ s as a function of $H$ for the same parameter set as in figure \ref{fig_no_6a}.  The lines correspond to the same quantities as in figure \ref{fig_no_1b}.


\paragraph{($p$, $\log |\xi_0|$) =($-1.5$, $82$):}
In this case, a positive real solution exits.  The solution, however, becomes much larger than that in SBBN, and this case is excluded based on the condition (4) (See section \ref{sec3}).

\begin{figure}[tbp]
\begin{center}
\includegraphics[width=8.0cm,clip]{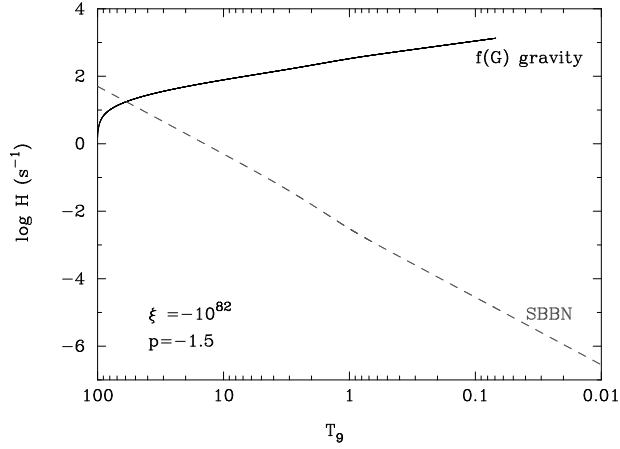}
\caption{Cosmic expansion rates $H$ as a function of $T_9$.  The solid line corresponds to the present $f(G)$ gravity model with ($p$, $\log |\xi_0|$, $f(G_\mathrm{i})$)=($-1.5$, $82$, $0$), while the dashed line corresponds to SBBN.  The difference between the two models becomes extremely large with decreasing temperature in the late epoch.\label{fig_no_7c}}
\end{center}
\end{figure}

Figure \ref{fig_no_7c} shows the cosmic expansion rates $H$ as a function of $T_9$.  The solid line corresponds to the present $f(G)$ gravity model with ($p$, $\log |\xi_0|$, $f(G_\mathrm{i})$)=($-1.5$, $82$, $0$), while the dashed line corresponds to SBBN.  When the expansion rate becomes $10^8$ times larger than the rate in SBBN, the calculation is terminated.  Although this case has a positive solution of $H$, this parameter region is safely excluded from the limits on elemental abundances.


\subsubsection{Effects on BBN}\label{sec5_2c}

\begin{figure}[tbp]
\begin{center}
\includegraphics[width=8.0cm,clip]{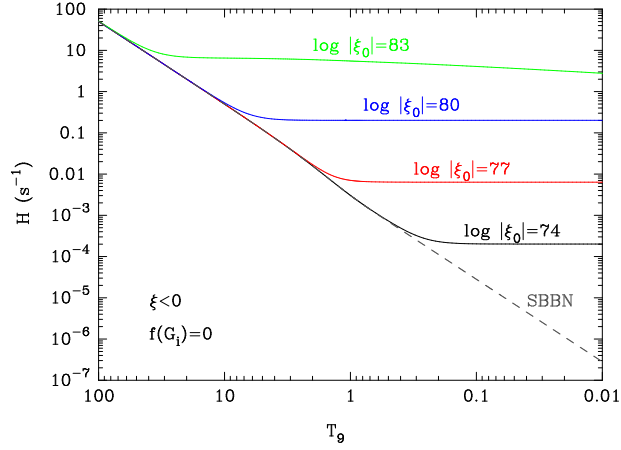}
\caption{Hubble expansion rate $H$ as a function of $T_9$ in SBBN (dashed line) and in the cases of $\log |\xi_0| =74$, $77$, $80$, and $83$, respectively, (solid lines).  It is assumed that the power-law index is $p=2$, and that the initial $f(G)$ value is zero, i.e, $f(G_\mathrm{i})=0$.  \label{fig_hubble_m1}}
\end{center}
\end{figure}

Figure \ref{fig_hubble_m1} shows the cosmic expansion rates $H$ (s$^{-1}$) as a function of $T_9$ in the SBBN model (dashed line) and four cases of $\log |\xi_0| =74$, $77$, $80$, and $83$, respectively, (solid lines) with fixed parameters of $p=2$ and $f(G_\mathrm{i})=0$ as an example case.  We note that the third and fifth terms in eq. \eqref{eq13} linearly scale as the first standard term, i.e., $-6H^2$, in the case of $p=2$ as long as the relation $H^\propto_\sim 1/t$ is not significantly broken (eqs. \eqref{eq27} and \eqref{eq29}).  In SBBN, the relation of $H \sim T_9^2$ is realized since the energy density is dominated by radiation in the BBN epoch.


\paragraph{Standard BBN:}

\begin{figure}[tbp]
\begin{center}
\includegraphics[width=8.0cm,clip]{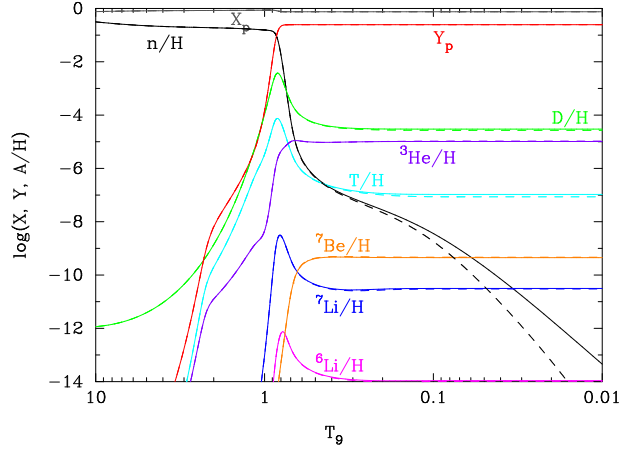}
\caption{Mass fractions of H and $^4$He ($X_{\rm p}$ and $Y_{\rm p}$, respectively) and number ratios of other nuclides relative to H as a function of $T_9$. Solid lines show the abundances in the $f(G)$ gravity model with the parameters ($p$, $\log |\xi_0|$, $[f(G)/(\xi G)]_\mathrm{i}$)=($2$, $74$, $0$), while dashed lines show those in SBBN.  \label{fig_m1_1}}
\end{center}
\end{figure}

Figure \ref{fig_m1_1} shows evolutions of elemental abundances as a function of $T_9$ in the SBBN model (dashed lines) and the $f(G)$ model with parameters ($p$, $\log |\xi_0|$, $[f(G)/(\xi G)]_\mathrm{i}$)=($2$, $74$, $0$) (solid lines).  The quantities $X_{\rm p}$ and $Y_{\rm p}$ are mass fractions of $^1$H and $^4$He, respectively.  Other lines show the number ratios of nuclides to hydrogen, $^1$H.


Firstly, we review important reactions in SBBN model.
At $T_9 \gtrsim 10$, by efficient weak reactions, the neutron-to-proton ratio is the equilibrium value of $(n/p)_{\rm eq}=\exp(-Q/T)$ with $Q=m_n -m_p =1.293$ MeV the mass difference.
After the weak reaction freeze-out at $T_9 \sim 10$, the $\beta$-decay of the neutron is the dominant reaction for neutron destruction.  When the temperature decreases to $T_9 \lesssim 1$, the neutron is destroyed mainly via $^1$H($n$, $\gamma$)$^2$H.
D is predominantly produced via $^1$H($n,\gamma$)$^2$H, and destroyed via $^2$H($d$, $n$)$^3$He and $^2$H($d$, $p$)$^3$H.
$^3$H is produced via $^2$H($d,p$)$^3$H and destroyed via $^3$H($d,n$)$^4$He.
$^3$He is produced via $^2$H($d,n$)$^3$He and destroyed via $^3$He($n,p$)$^3$H.
The primordial $^3$H abundance is the sum of abundances of $^3$H and $^3$He produced during the BBN.  Long after the BBN, $^3$He nuclei $\beta$-decay into $^3$H nuclei with the half-life of 12.32 y.  The final abundance of $^3$He is larger than that of $^3$H by about two orders of magnitude in SBBN.  Therefore, the primordial $^3$H abundance predominantly reflects the larger $^3$He abundance during BBN.

$^7$Li is produced via $^4$He($t,\gamma$)$^7$Li and destroyed via $^7$Li($p,\alpha$)$^4$He.
$^7$Be is produced via $^4$He($^3$He$,\gamma$)$^7$Be and destroyed via $^7$Be$(n,p$)$^7$Li.
The primordial $^7$Li abundance is the sum of abundances of $^7$Li and $^7$Be produced during the BBN.  Long after the BBN, $^7$Be nuclei recombine with electrons and are transformed to $^7$Li nuclei via the electron capture process.  $^7$Be is produced more than $^7$Li in SBBN.  The primordial $^7$Li abundance then predominantly reflects the larger $^7$Be abundance during BBN.
$^6$Li is produced via $^4$He($d,\gamma$)$^6$Li and destroyed via $^6$Li$(p,\alpha$)$^3$He.

Results of BBN in the $f(G)$ model are compared with those in SBBN below.

\paragraph{$\log |\xi_0|=74$:}

The expansion rate becomes larger than that in SBBN at $T_9 \lesssim 0.4$ (figure \ref{fig_hubble_m1}). 
The temperature then decreases in a time scale shorter than in SBBN after the temperature comes down to this critical value.
Because of the faster expansion, the deuterium destruction is less efficient, and the D/H abundance is higher (figure \ref{fig_m1_1}).

\paragraph{$\log |\xi_0|=77$:}

\begin{figure}[tbp]
\begin{center}
\includegraphics[width=8.0cm,clip]{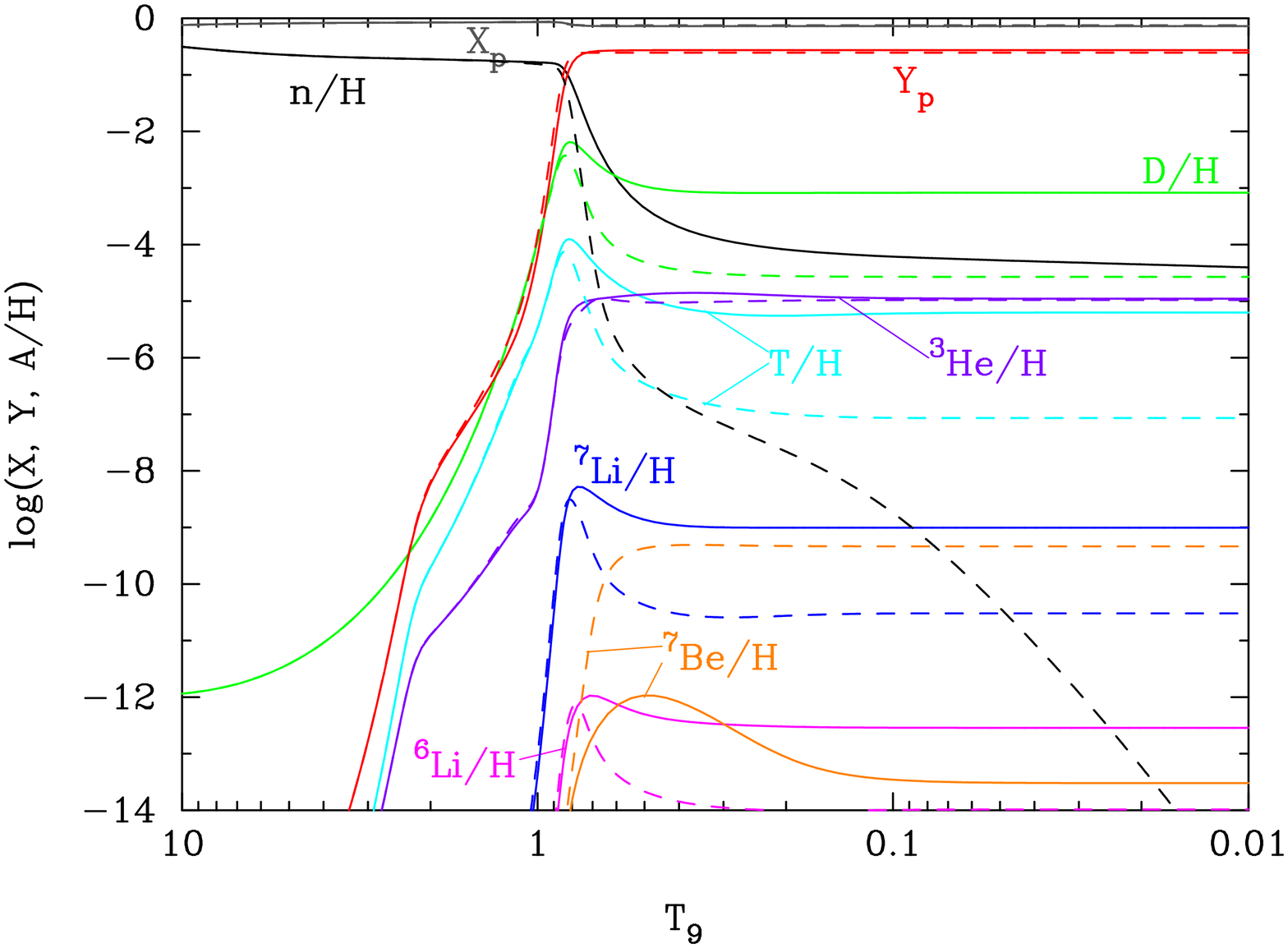}
\caption{Same as in figure \ref{fig_m1_1}, but solid lines correspond to the $f(G)$ gravity model with the parameters ($p$, $\log |\xi_0|$, $[f(G)/(\xi G)]_\mathrm{i}$)=($2$, $77$, $0$).  \label{fig_m1_2}}
\end{center}
\end{figure}

Figure \ref{fig_m1_2} shows elemental abundances as a function of $T_9$, similarly to figure \ref{fig_m1_1}, in the case of parameters ($p$, $\log |\xi_0|$, $[f(G)/(\xi G)]_\mathrm{i}$)=($2$, $77$, $0$) (solid lines).  
The expansion rate becomes larger than that in SBBN at $T_9 \lesssim 2$ (figure \ref{fig_hubble_m1}).
The cosmic time at a given temperature is then shorter.  As a result, the $n$/$p$ ratio is higher at the $^4$He synthesis temperature of $T_9 \sim 1$, and $^4$He abundance is then slightly higher.  We note that the $^4$He synthesis occurs at slightly lower temperature than in SBBN because of the shorter expansion time scale for a given temperature.
The destructions of D and $^3$H freeze out earlier than in SBBN.  The destructions are, therefore, less efficient, and the abundances D/H and $^3$H/H are higher.
The production and destruction rates of $^3$He is higher because of higher D abundance.  As a result, the abundance $^3$He/H is slightly larger.
Destructions of $^7$Li and $^6$Li freeze out earlier than in SBBN.  The destructions are then less efficient, and the abundances $^7$Li/H and $^6$Li/H are higher.
The $^7$Be abundance is very small because of efficient destruction by abundant neutrons via $^7$Be($n$, $p$)$^7$Li.


\paragraph{$\log |\xi_0|=80$:}\label{caseb11}

\begin{figure}[tbp]
\begin{center}
\includegraphics[width=8.0cm,clip]{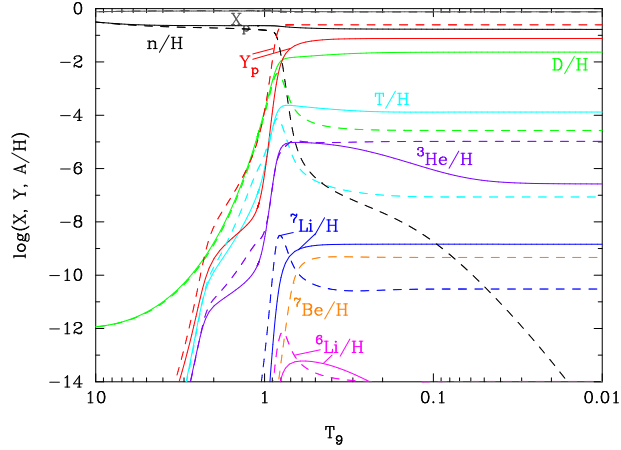}
\caption{Same as in figure \ref{fig_m1_1}, but solid lines correspond to the $f(G)$ gravity model with the parameters ($p$, $\log |\xi_0|$, $[f(G)/(\xi G)]_\mathrm{i}$)=($2$, $80$, $0$).  \label{fig_m1_3}}
\end{center}
\end{figure}

Figure \ref{fig_m1_3} shows elemental abundances as a function of $T_9$, similarly to figure \ref{fig_m1_1}, in the case of parameters ($p$, $\log |\xi_0|$, $[f(G)/(\xi G)]_\mathrm{i}$)=($2$, $80$, $0$) (solid lines).  
The expansion rate becomes larger than that in SBBN at $T_9 \lesssim 10$.
The $^4$He production is incomplete, and the $^4$He synthesis occurs at lower temperature than in SBBN.  The $^4$He abundance is then lower.
The destructions of D and $^3$H are less efficient, and the abundances D/H and $^3$H/H are higher.
The neutron abundance is very high since the expansion time scale is short compared to the neutron $\beta$-decay half-life.
$^3$He is then effectively converted to $^3$H by neutrons via $^3$He($n$, $p$)$^3$H.  The abundance $^3$He/H is, therefore, smaller.
The nuclei $^7$Li and $^6$Li are produced at lower temperature.  The destructions are then less efficient, and the abundances $^7$Li/H and $^6$Li/H are higher.
The $^7$Be abundance is very small because of efficient destruction via $^7$Be($n$, $p$)$^7$Li.


\paragraph{$\log |\xi_0|=83$:}

\begin{figure}[tbp]
\begin{center}
\includegraphics[width=8.0cm,clip]{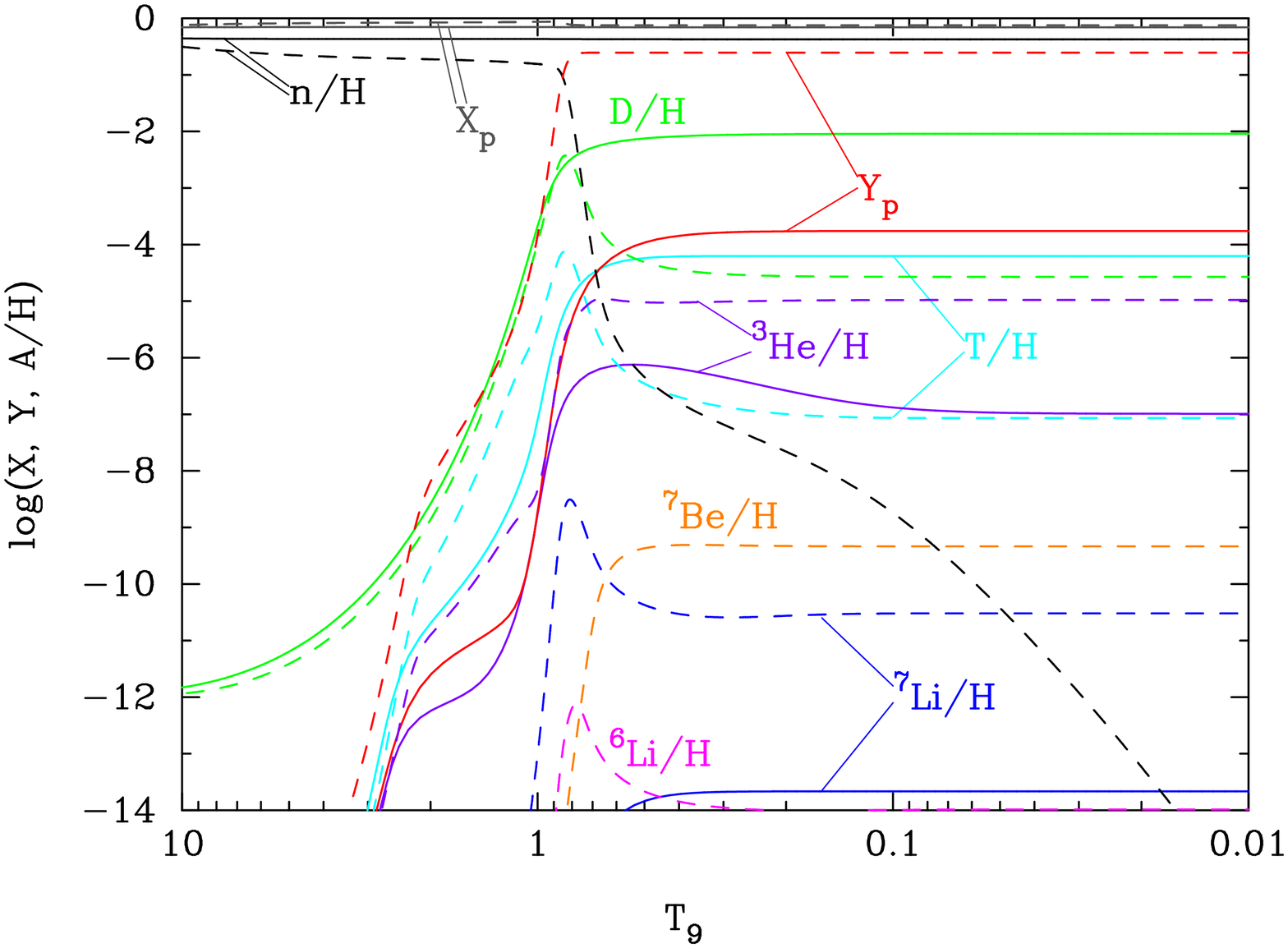}
\caption{Same as in figure \ref{fig_m1_1}, but solid lines correspond to the $f(G)$ gravity model with the parameters ($p$, $\log |\xi_0|$, $[f(G)/(\xi G)]_\mathrm{i}$)=($2$, $83$, $0$).  \label{fig_m1_4}}
\end{center}
\end{figure}

Figure \ref{fig_m1_4} shows elemental abundances as a function of $T_9$, similarly to figure \ref{fig_m1_1}, in the case of parameters ($p$, $\log |\xi_0|$, $[f(G)/(\xi G)]_\mathrm{i}$)=($2$, $83$, $0$) (solid lines).  
The expansion rate becomes larger than that in SBBN at $T_9 \lesssim 50$.
The $^4$He production is incomplete, and the $^4$He synthesis occurs at lower temperature than in SBBN.  The $^4$He abundance is then much lower. ($Y_{\rm p} \sim 10^{-4}$).
The destructions of D and $^3$H are less efficient.  Their productions are, however, also less efficient.  Therefore, the abundances D/H and $^3$H/H are higher than in SBBN although smaller than in case (d).
The neutron abundance is very high.
$^3$He is then effectively converted to $^3$H via $^3$He($n$, $p$)$^3$H so that its abundance is smaller.
The nuclei $^7$Li and $^7$Be are produced at lower temperature.  Their abundances are, however, small because of less abundances of $^4$He, $^3$H, and $^3$He.  The $^7$Li destruction is less efficient.  As a result, the abundance $^7$Li/H is smaller.
The $^6$Li abundance is very small because of a smaller abundance of $^4$He.
In addition, the $^7$Be abundance is effectively reduced by efficient destruction via $^7$Be($n$, $p$)$^7$Li.


\subsection{Case of $\xi_0 >0$ and $p=2$}\label{sec5_3}

\subsubsection{Parameter search}\label{sec5_3a}

\begin{figure}[tbp]
\begin{center}
\includegraphics[width=8.0cm,clip]{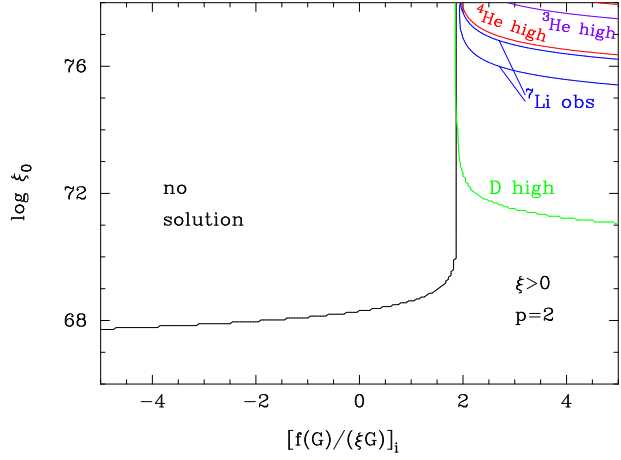}
\caption{Same boundary (black line) and contours of calculated light element abundances (colored lines) as in figure \ref{fig2}, but in the parameter plane of ($[f(G)/(\xi G)]_\mathrm{i}$, $\xi_0$) for $\xi_0>0$.  For this figure, the power-law index is fixed as $p=2$.  \label{fig3}}
\end{center}
\end{figure}

Figure \ref{fig3} shows the same boundary (black line) and contours of element abundances (colored lines) as in figure \ref{fig2}, but in the parameter plane of ($[f(G)/(\xi G)]_\mathrm{i}$, $\xi_0$) for $\xi_0>0$.  The power-law index is fixed as $p=2$.  When the condition $[f(G)/(\xi G)]_\mathrm{i} <2$ is satisfied initially, the constraint of $\log \xi_0 \lesssim 68$ is derived from the requirement for the continuous existence of solution of $H$ during the BBN.  In the case of $[f(G)/(\xi G)]_\mathrm{i} >2$, on the other hand, the constraint from the existence of solution disappears, and constraints from observational light element abundances are important.  The reasons are described as follows.


We suppose that the terms of modified gravity do not affect the cosmic expansion rate at the initial time very much.   Then, because of $G_\mathrm{i}<0$ (eqs. \eqref{eq6} and \eqref{eq24}) and $\xi_\mathrm{i}>0$, the inequality $\xi_\mathrm{i} G_\mathrm{i} <0$ is satisfied.  The region of $f(G_\mathrm{i})/(\xi_\mathrm{i} G_\mathrm{i} )<0$, therefore, corresponds to $f(G_\mathrm{i})>0$.  As the value of $f(G_\mathrm{i})>0$ is larger, the negative-$\Lambda$-like effect of the $f(G)$ term is stronger.  As a result, smaller values of $\xi_0$ is allowed for smaller values of $f(G_\mathrm{i})/(\xi_\mathrm{i} G_\mathrm{i} )$ by the consideration of the existence of solution.  On the other hand, the region of $f(G_\mathrm{i})/(\xi_\mathrm{i} G_\mathrm{i} )>0$ corresponds to $f(G_\mathrm{i})<0$.  As the value of $f(G_\mathrm{i})$ is smaller, it is more difficult to lose a proper solution of $H$.  Therefore, the constraint from the existence of solution is weaker.  

Using physical quantities in standard model (as in eqs. \eqref{eq26}--\eqref{eq29}), a somewhat detailed equation of the $[f(G) -f(G_\mathrm{i})]$ is derived as
\begin{equation}
-\kappa^2 \left[f(G) -f(G_\mathrm{i}) \right] \sim \left\{
\begin{array}{ll}
\kappa^2 \frac{4}{p-4} \xi G & (\mathrm{for}~p \gtrsim 4) \\
\kappa^2 \frac{4}{4-p} \xi_\mathrm{i} G_\mathrm{i} & (\mathrm{for}~p \lesssim 4). \\
\end{array}
\right.
\label{eq31}
\end{equation}

For example, when the power-law index is $p=2$ as in the present case, it follows
\begin{equation}
\label{eq32}
-\kappa^2 \left[f(G) -f(G_\mathrm{i}) \right] \sim
 2\kappa^2 \xi_\mathrm{i} G_\mathrm{i} <0.
\end{equation}
Then, when the inequality $f(G_\mathrm{i}) \leq 2 \xi_\mathrm{i} G_\mathrm{i} $ or $f(G_\mathrm{i})/(\xi_\mathrm{i} G_\mathrm{i}) \geq 2$ is satisfied, the condition $- \kappa^2 f(G) \sim \kappa^2 [2 \xi_\mathrm{i} G_\mathrm{i} -f(G_\mathrm{i})] \geq 0$ always holds.  The $- \kappa^2 f(G)$ term, therefore, operates similarly to a positive $\Lambda$.  In this case, the real positive solution of $H$ never disappears, and the `no solution' region is absent in figure \ref{fig3}. In the region of $[f(G)/(\xi G)]_\mathrm{i} \geq 2$, when the amplitude $\xi_0$ is larger, the Hubble expansion rate is larger because of the $f(G)$ term.  Therefore, effects on the light element abundances are larger.  Observational limits from elemental abundances then appear in the right upper corner in figure \ref{fig3}.  For larger values of $[f(G)/(\xi G)]_\mathrm{i}$ or smaller values of $f(G_\mathrm{i})$, the increase of the cosmic expansion rate from that in SBBN is larger.  Therefore, constraints from elemental abundances are severer.  This fact explains the downward-sloping abundance contours in the region of $[f(G)/(\xi G)]_\mathrm{i} \geq 2$.

\paragraph{$p \lesssim 4$:}

In general, a real positive solution of $H$ exists when the following condition is satisfied,
\begin{eqnarray}
\lim_{H \to +0} g(H; \xi, \rho) &=& \lim_{H \to +0} 2 \kappa^2 \rho -6 H^2 - \kappa^2 f(G) -12\kappa^4 \left( \rho +p \right) H^2 \xi \nonumber \\
&=& \lim_{H \to +0} 2 \kappa^2 \rho - \kappa^2 f(G) -6 \left[1+ 2\kappa^4 \left( \rho +p \right) \xi  \right] H^2  \nonumber \\
&=&
2 \kappa^2 \rho - \kappa^2 f(G)
>0. \label{eq33}
\end{eqnarray}
We note that in the limit as $H$ approaches 0 from the right, the third term in eq. \eqref{eq13} ($\propto H^3$) is negligible compared to the $-6H^2$ term, and that the GB term has the limit value of $\lim_{H \to +0} G =-12 \kappa^2 (\rho +p) H^2$ (eqs. \eqref{eq6} and \eqref{eq14}).  Equation \eqref{eq33} is satisfied when the following condition holds,
\begin{eqnarray}
f(G) \sim f(G_\mathrm{i}) -\frac{4}{4-p} \xi_\mathrm{i} G_\mathrm{i}
&<&0 \nonumber \\
\Longrightarrow ~~~~~\frac{f(G_\mathrm{i})}{\xi_\mathrm{i} G_\mathrm{i}} & \gtrsim & \frac{4}{4-p},
\label{eq34}
\end{eqnarray}
where
eq. \eqref{eq31} was used in the first line.

\paragraph{$p \gtrsim 4$:}

A real positive solution of $H$ exists when eq. \eqref{eq33} is satisfied.  In this case, however, eq. \eqref{eq31} is approximately true only when the cosmic expansion rate is not much different from that in the standard model.  As the amplitude $|f(G)|$ becomes relatively large with time, the expansion rate can be reduced.  We define this epoch as the deviation time $t_\mathrm{dev}$.  Then, $G \propto H^2 (\dot{H} +H^2)$ becomes very small after the deviation time.  The contribution to the integration in the $f(G)$ function (eq. \eqref{eq16}) from the time after $t_\mathrm{dev}$ is then small.  As a result, the quantity of eq. \eqref{eq16} is mainly given by the value in the standard model at $t=t_\mathrm{dev}$.  Equation \eqref{eq29}, therefore, leads (cf. eq. \eqref{eq34}) to the condition,
\begin{eqnarray}
f(G) \sim f(G_\mathrm{i}) -\frac{4}{p-4} \left( \xi G \right)_{\rm dev}
&<&0 \nonumber \\
\Longrightarrow ~~~~~\frac{f(G_\mathrm{i})}{\left( \xi G \right)_{\rm dev}} & \gtrsim & \frac{4}{p-4},
\label{eq35}
\end{eqnarray}
where
$( \xi G )_{\rm dev}$ is the value of $\xi G$ at the deviation time.  Before the deviation time, since the quantity $(\xi G)$ has a simple scaling (eq. \eqref{eq29}), it follows that
\begin{equation}
\label{eq36}
\left( \xi G \right) = \xi_\mathrm{i} G_\mathrm{i} \left( t/ t_\mathrm{i} \right)^{p-4}.
\end{equation}
The following condition is then found,
\begin{equation}
\label{eq37}
\frac{f(G_\mathrm{i})}{\left( \xi_\mathrm{i} G_\mathrm{i} \right)} \gtrsim \frac{4}{p-4} \left( \frac{t_\mathrm{dev}}{t_\mathrm{i}} \right)^{p-4}.
\end{equation}

\subsubsection{Effects on BBN}\label{sec5_3b}

\begin{figure}[tbp]
\begin{center}
\includegraphics[width=8.0cm,clip]{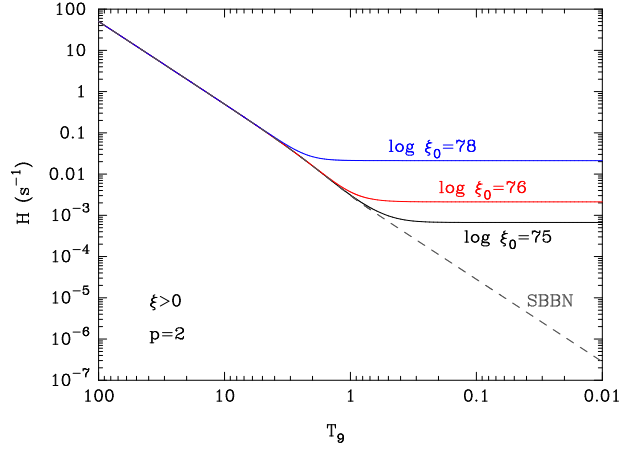}
\caption{Hubble expansion rate $H$ as a function of $T_9$ in SBBN (dashed line) and in the cases of $\log \xi_0 =75$, $76$, and $78$, respectively, with fixed parameters of $p=2$ and $[f(G)/(\xi G)]_\mathrm{i}=4$ (solid lines).  \label{fig_hubble_p2}}
\end{center}
\end{figure}

Figure \ref{fig_hubble_p2} shows the cosmic expansion rates $H$ (s$^{-1}$) as a function of $T_9$ in the SBBN model (dashed line) and three cases of $\log \xi_0 =75$, $76$, and $78$, respectively, (solid lines) with fixed parameters of $p=2$ and $[f(G)/(\xi G)]_\mathrm{i}=4$.


\paragraph{($\log \xi_0$, $[f(G)/(\xi G)]_\mathrm{i}$)=($75$, $4$):}\label{casea21}

\begin{figure}[tbp]
\begin{center}
\includegraphics[width=8.0cm,clip]{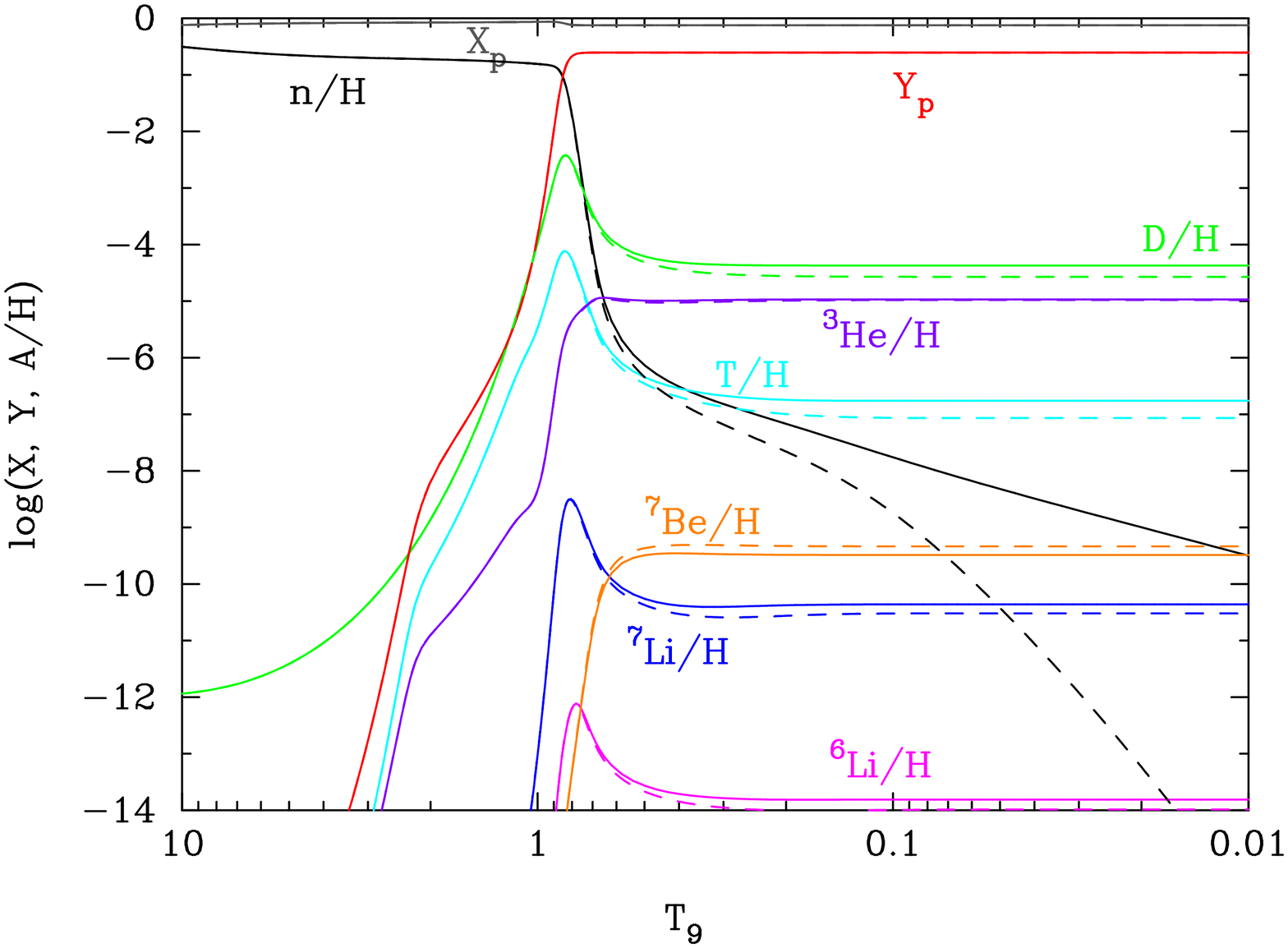}
\caption{Same as in figure \ref{fig_m1_1}, but solid lines correspond to the $f(G)$ gravity model with the parameters ($p$, $\log \xi_0$, $[f(G)/(\xi G)]_\mathrm{i}$)=($2$, $75$, $4$).  \label{fig_p2_1}}
\end{center}
\end{figure}

Figure \ref{fig_p2_1} shows elemental abundances as a function of $T_9$, similarly to figure \ref{fig_m1_1}, in the case of parameters ($p$, $\log \xi_0$, $[f(G)/(\xi G)]_\mathrm{i}$)=($2$, $75$, $4$) (solid lines).  
The expansion rate becomes larger than that in SBBN at $T_9 \lesssim 0.7$ (figure \ref{fig_hubble_p2}). 
Deuterium destruction is then less efficient, and the final abundance D/H is higher.
$^3$H abundance is also somewhat larger.  The neutron abundance remains to be high since the temperature decreases effectively without the change of cosmic time after the expansion rate deviates from that in SBBN. 
Destructions of $^7$Li and $^6$Li, and $^7$Be production freeze out earlier than in SBBN.  Therefore, the abundances $^7$Li/H and $^6$Li/H are higher while the $^7$Be/H abundance is lower.


\paragraph{($\log \xi_0$, $[f(G)/(\xi G)]_\mathrm{i}$)=($76$, $4$):}

\begin{figure}[tbp]
\begin{center}
\includegraphics[width=8.0cm,clip]{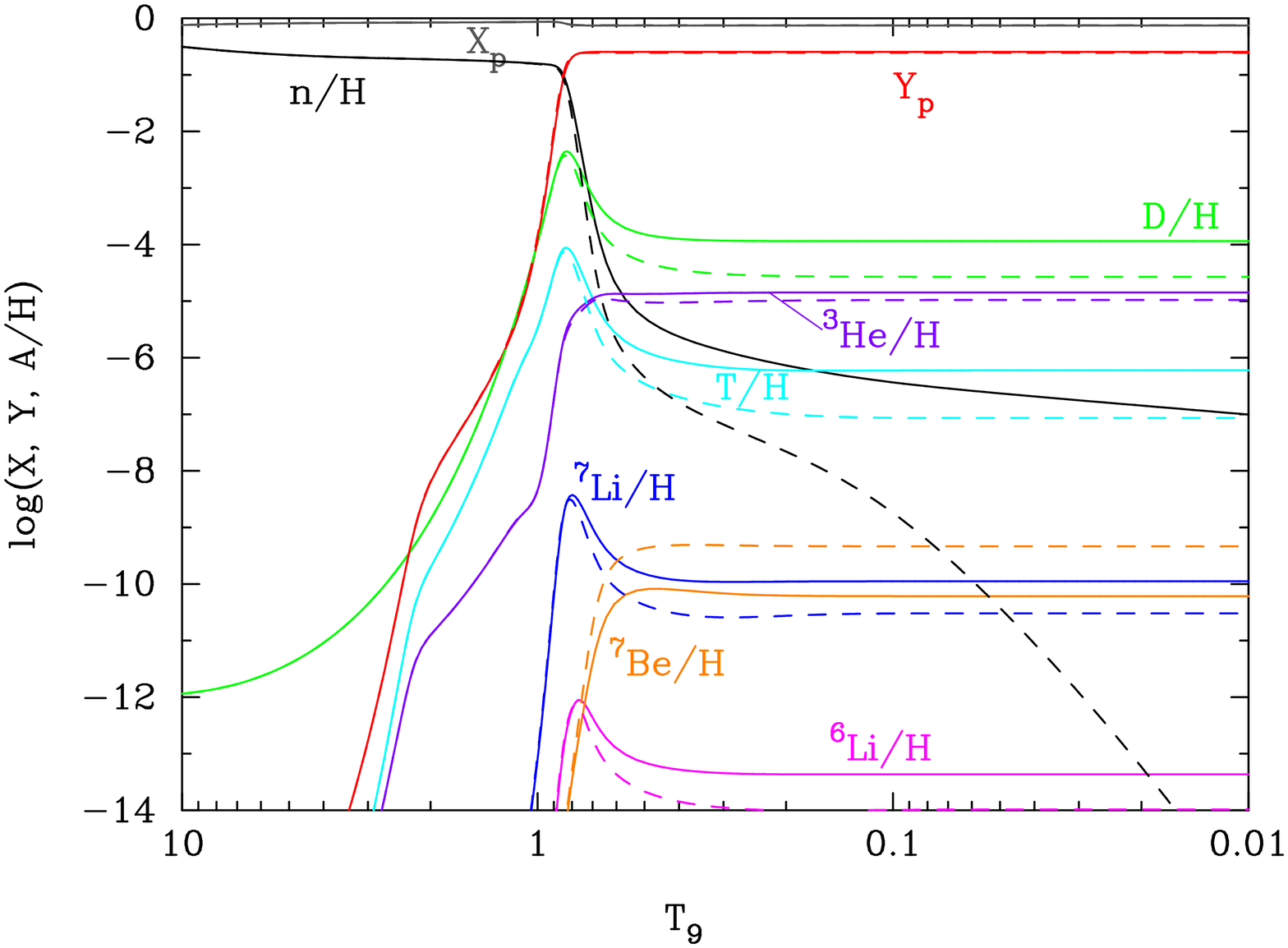}
\caption{Same as in figure \ref{fig_m1_1}, but solid lines correspond to the $f(G)$ gravity model with the parameters ($p$, $\log \xi_0$, $[f(G)/(\xi G)]_\mathrm{i}$)=($2$, $76$, $4$).  \label{fig_p2_2}}
\end{center}
\end{figure}

Figure \ref{fig_p2_2} shows elemental abundances as a function of $T_9$, similarly to figure \ref{fig_m1_1}, in the case of parameters ($p$, $\log \xi_0$, $[f(G)/(\xi G)]_\mathrm{i}$)=($2$, $76$, $4$) (solid lines).  
The expansion rate becomes larger than that in SBBN at $T_9 \lesssim 1$ (figure \ref{fig_hubble_p2}). 
In this case, destructions of D, $^3$H, and $^3$He are less efficient, and the final abundances of D/H, $^3$H/H, and $^3$He/H are higher than those in SBBN.
Also, destructions of $^7$Li and $^6$Li, and $^7$Be production freeze out earlier than in SBBN, similarly to the case (a).
It then results in significantly higher $^7$Li/H and $^6$Li/H and lower $^7$Be/H values.


\paragraph{($\log \xi_0$, $[f(G)/(\xi G)]_\mathrm{i}$)=($78$, $4$):}

\begin{figure}[tbp]
\begin{center}
\includegraphics[width=8.0cm,clip]{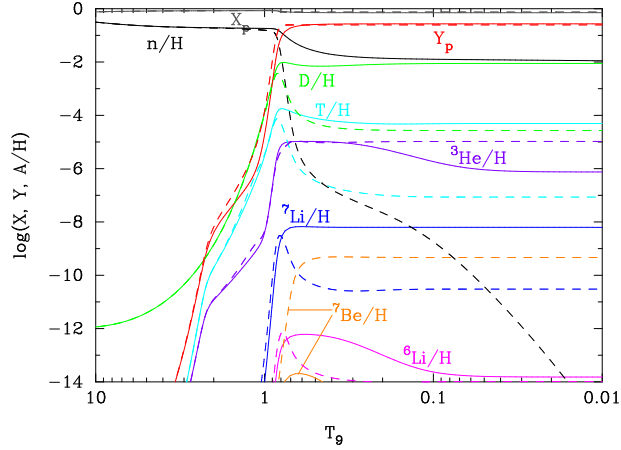}
\caption{Same as in figure \ref{fig_m1_1}, but solid lines correspond to the $f(G)$ gravity model with the parameters ($p$, $\log \xi_0$, $[f(G)/(\xi G)]_\mathrm{i}$)=($2$, $78$, $4$).  \label{fig_p2_3}}
\end{center}
\end{figure}

Figure \ref{fig_p2_3} shows elemental abundances as a function of $T_9$, similarly to figure \ref{fig_m1_1}, in the case of parameters ($p$, $\log \xi_0$, $[f(G)/(\xi G)]_\mathrm{i}$)=($2$, $78$, $4$) (solid lines).  
The expansion rate becomes larger than that in SBBN at $T_9 \lesssim 3$. 
The cosmic time at a given temperature is then shorter.  As a result, the $n$/$p$ ratio is higher at the $^4$He synthesis temperature of $T_9 \sim 1$, and $^4$He abundance is then slightly higher.  We note that the $^4$He synthesis occurs at slightly lower temperature than in SBBN because of the shorter expansion time scale for a given temperature.
The destructions of D and $^3$H are less efficient, and the abundances D/H and $^3$H/H are higher.
The neutron abundance remains to be high ($n$/H $\sim 10^{-2}$).
$^3$He is then effectively converted to $^3$H by abundant neutrons via $^3$He($n$, $p$)$^3$H.  The abundance $^3$He/H is, therefore, smaller.
The nuclei $^7$Li and $^6$Li are produced at slightly lower temperature than in SBBN.
Since the $^7$Li destruction is less efficient at lower temperature, the abundance $^7$Li/H is higher.
The $^7$Be abundance is very small because of efficient destruction by neutrons via $^7$Be($n$, $p$)$^7$Li.
The abundance $^6$Li/H is higher than in SBBN.  It is, however, reduced at $T_9 \sim 0.5-0.1$ via $^6$Li($n$, $\alpha$)$^3$H, which is never important in SBBN.


\subsection{Case of $\xi_0 <0$ and $p=2$}\label{sec5_4}

\subsubsection{Parameter search}\label{sec5_4a}

\begin{figure}[tbp]
\begin{center}
\includegraphics[width=8.0cm,clip]{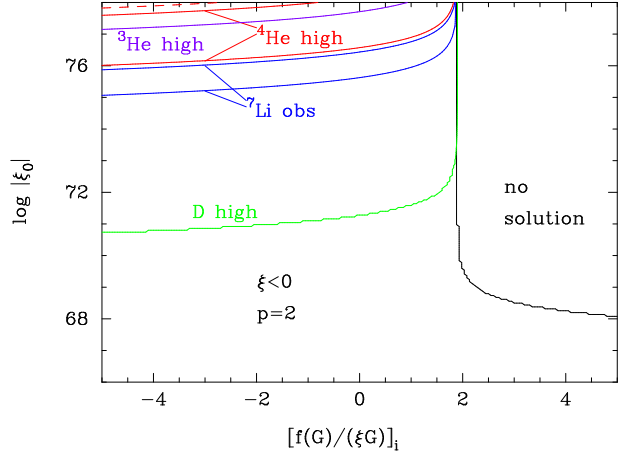}
\caption{Same as in figure \ref{fig3}, but in the parameter plane of ($[f(G)/(\xi G)]_\mathrm{i}$, $|\xi_0|$) for $\xi_0<0$.  \label{fig4}}
\end{center}
\end{figure}

Figure \ref{fig4} shows the same contours as in figure \ref{fig3} but for $\xi_0<0$.  Parameter regions in figures \ref{fig3} and \ref{fig4} looks mirror symmetric with respect to the $f(G_\mathrm{i})/(\xi_\mathrm{i} G_\mathrm{i}) =2$ line.  The solution boundary and abundance contours are explained as follows.


We again suppose that the terms of modified gravity do not affect the cosmic expansion rate at the initial time very much.   The inequalities $G_\mathrm{i}<0$ and $\xi_\mathrm{i}<0$ are satisfied, and the inequality $\xi_\mathrm{i} G_\mathrm{i} >0$ holds.  The region of $f(G_\mathrm{i})/(\xi_\mathrm{i} G_\mathrm{i})>0$, therefore, corresponds to $f(G_\mathrm{i})>0$.  Larger values of $f(G_\mathrm{i})$ then lead to stronger effects of the negative-$\Lambda$-like $f(G)$ term.  Smaller values of $\xi_0$ are then allowed for larger values of $f(G_\mathrm{i})/(\xi_\mathrm{i} G_\mathrm{i})$ by the consideration of the existence of solution.  On the other hand, the region of $f(G_\mathrm{i})/(\xi_\mathrm{i} G_\mathrm{i})<0$ corresponds to $f(G_\mathrm{i})<0$.  When $f(G_\mathrm{i})$ is smaller, it is more difficult to lose a proper solution of $H$, and the constraint from the existence of solution is weaker.  However, the cosmic expansion rate is increased more, and light element abundances are affected more.  As a result, constraints from the observational limits on the abundances are stronger, which can be seen from the upward-sloping abundance contours for $[f(G)/(\xi G)]_\mathrm{i}<2$.

Similarly to the case of $\xi_0>0$, eqs. \eqref{eq31} and \eqref{eq36} are true before the deviation time.  
\paragraph{$p \lesssim 4$:}

The condition for the existence of proper solution of $H$ is given by $f(G)\lesssim 0$.  This is satisfied if
\begin{equation}
\label{eq38}
\frac{f(G_\mathrm{i})}{\xi_\mathrm{i} G_\mathrm{i} } \lesssim \frac{4}{4-p}.
\end{equation}

\paragraph{$p \gtrsim 4$:}

The condition for the existence of the solution $H$ is satisfied when the following condition holds,
\begin{equation}
\label{eq39}
\frac{f(G_\mathrm{i})}{\left( \xi_\mathrm{i} G_\mathrm{i} \right)} \lesssim \frac{4}{p-4} \left( \frac{t_\mathrm{dev}}{t_\mathrm{i}} \right)^{p-4}.
\end{equation}

For example, for the power-law index $p=2$, the condition for a proper solution is given (eq. \eqref{eq38}) by $f(G_\mathrm{i})/(\xi_\mathrm{i} G_\mathrm{i}) \lesssim 2$.  If this condition is not met, i.e., $f(G_\mathrm{i})/(\xi_\mathrm{i} G_\mathrm{i}) \gtrsim 2$, the term $f(G)>0$ induces a negative-$\Lambda$ effect, and the solution of $H$ can be lost.

\subsubsection{Effects on BBN}\label{sec5_4b}

\begin{figure}[tbp]
\begin{center}
\includegraphics[width=8.0cm,clip]{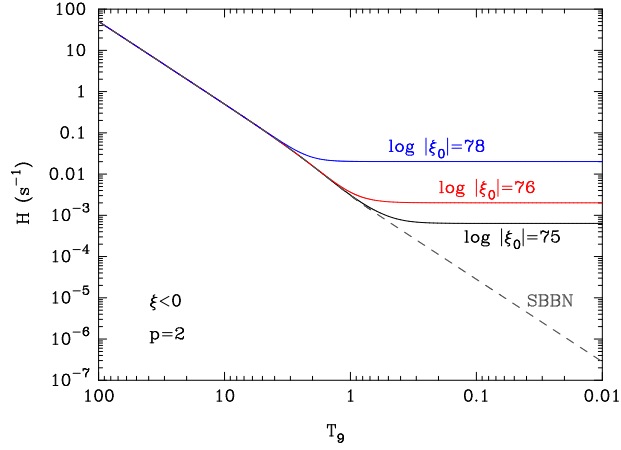}
\caption{Hubble expansion rate $H$ as a function of $T_9$ in SBBN (dashed line) and in the cases of $\log |\xi_0| =75$, $76$, and $78$, respectively, with fixed parameters of $p=2$ and $[f(G)/(\xi G)]_\mathrm{i}=4$ (solid lines).  \label{fig_hubble_m2}}
\end{center}
\end{figure}

Figure \ref{fig_hubble_m2} shows the cosmic expansion rates $H$ (s$^{-1}$) as a function of $T_9$ in the SBBN model (dashed line) and three cases of $\log |\xi_0| =75$, $76$, and $78$, respectively, with fixed parameters of $p=2$ and $[f(G)/(\xi G)]_\mathrm{i}=4$ (solid lines).


Trends of nucleosynthesis are very similar to those in section \ref{sec5_3b}.

\paragraph{($\log |\xi_0|$, $[f(G)/(\xi G)]_\mathrm{i}$)=($75$, $4$):}

\begin{figure}[tbp]
\begin{center}
\includegraphics[width=8.0cm,clip]{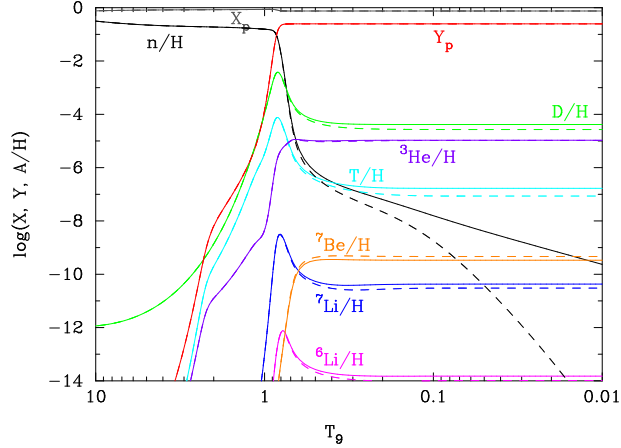}
\caption{Same as in figure \ref{fig_m1_1}, but solid lines correspond to the $f(G)$ gravity model with the parameters ($p$, $\log |\xi_0|$, $[f(G)/(\xi G)]_\mathrm{i}$)=($2$, $75$, $4$).  \label{fig_m2_1}}
\end{center}
\end{figure}

Figure \ref{fig_m2_1} shows elemental abundances as a function of $T_9$, similarly to figure \ref{fig_m1_1}, in the case of parameters ($p$, $\log |\xi_0|$, $[f(G)/(\xi G)]_\mathrm{i}$)=($2$, $75$, $4$) (solid lines).  
The expansion rate becomes larger than that in SBBN at $T_9 \lesssim 0.7$ (figure \ref{fig_hubble_m2}).
For the same reasons as in case (a) in section \ref{sec5_3b}, the abundances $n$/H, D/H, $^3$He/H, $^7$Li/H, and $^6$Li/H are higher while $^7$Be/H is lower than in SBBN.


\paragraph{($\log |\xi_0|$, $[f(G)/(\xi G)]_\mathrm{i}$)=($76$, $4$):}

\begin{figure}[tbp]
\begin{center}
\includegraphics[width=8.0cm,clip]{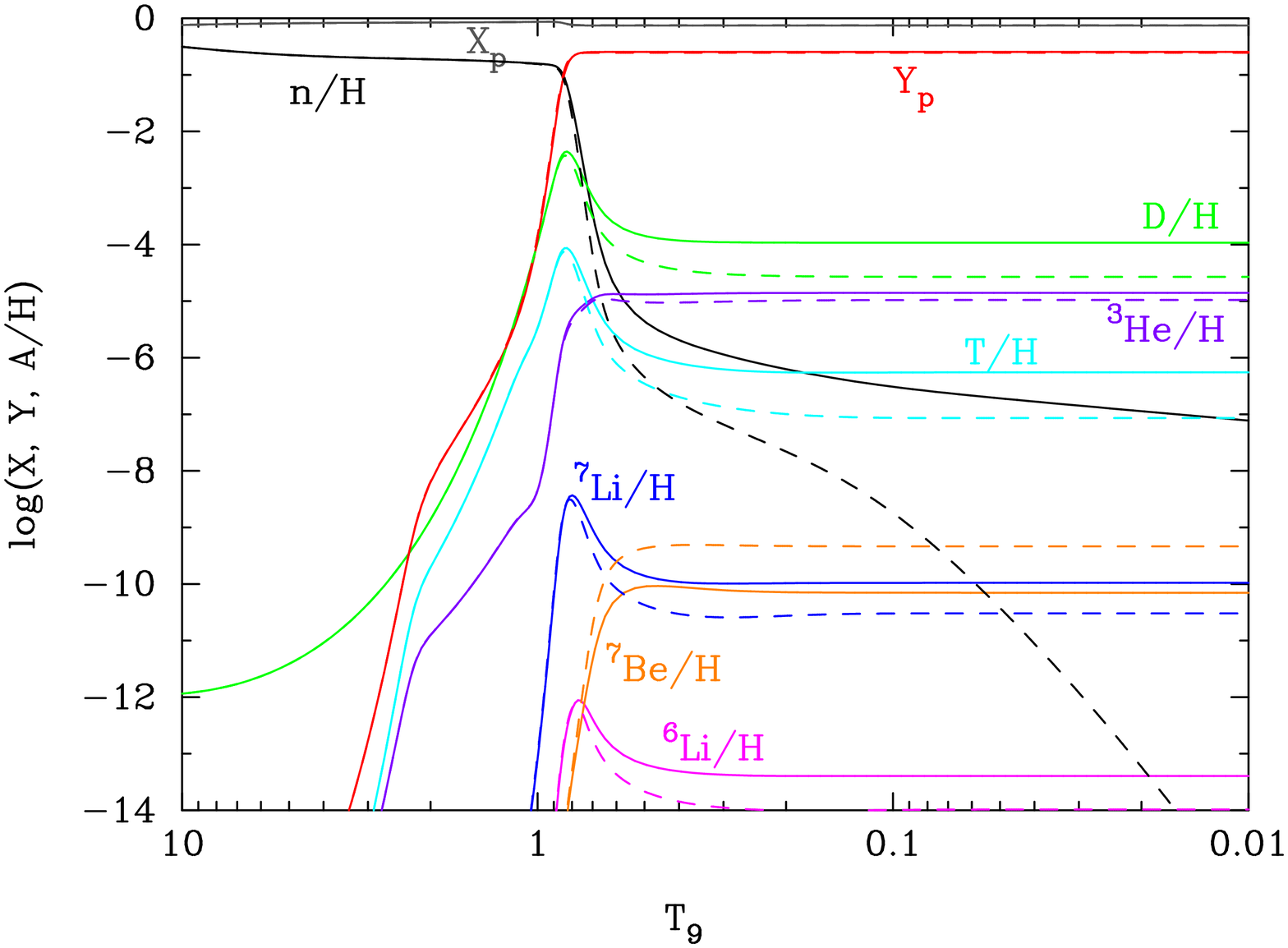}
\caption{Same as in figure \ref{fig_m1_1}, but solid lines correspond to the $f(G)$ gravity model with the parameters ($p$, $\log |\xi_0|$, $[f(G)/(\xi G)]_\mathrm{i}$)=($2$, $76$, $4$).  \label{fig_m2_2}}
\end{center}
\end{figure}

Figure \ref{fig_m2_2} shows elemental abundances as a function of $T_9$, similarly to figure \ref{fig_m1_1}, in the case of parameters ($p$, $\log |\xi_0|$, $[f(G)/(\xi G)]_\mathrm{i}$)=($2$, $76$, $4$) (solid lines).  
The expansion rate becomes larger than that in SBBN at $T_9 \lesssim 1$ (figure \ref{fig_hubble_m2}).
Although the trends of abundances are qualitatively the same as those in the case (a), differences in abundances from the SBBN are larger in case (b) because of larger difference in the expansion rate.


\paragraph{($\log |\xi_0|$, $[f(G)/(\xi G)]_\mathrm{i}$)=($78$, $4$):}

\begin{figure}[tbp]
\begin{center}
\includegraphics[width=8.0cm,clip]{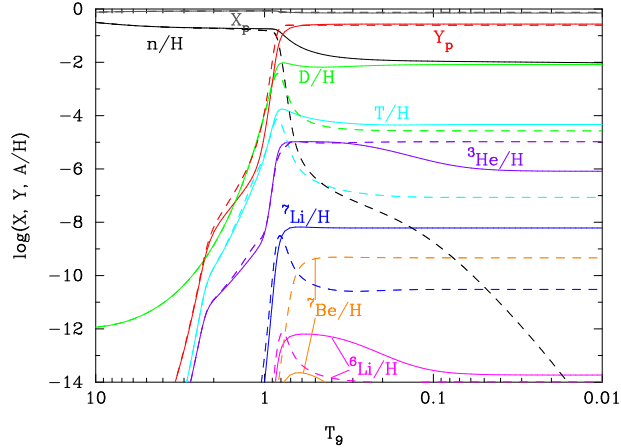}
\caption{Same as in figure \ref{fig_m1_1}, but solid lines correspond to the $f(G)$ gravity model with the parameters ($p$, $\log |\xi_0|$, $[f(G)/(\xi G)]_\mathrm{i}$)=($2$, $78$, $4$).  \label{fig_m2_3}}
\end{center}
\end{figure}

Figure \ref{fig_m2_3} shows elemental abundances as a function of $T_9$, similarly to figure \ref{fig_m1_1}, in the case of parameters ($p$, $\log |\xi_0|$, $[f(G)/(\xi G)]_\mathrm{i}$)=($2$, $78$, $4$) (solid lines).  
The expansion rate becomes larger than that in SBBN at $T_9 \lesssim 3$ (figure \ref{fig_hubble_m2}).
Because of the faster expansion, the $n$/$p$ ratio is higher at $T_9 \sim 1$, and $^4$He abundance is slightly higher.  
The $^4$He synthesis occurs at slightly lower temperature than in SBBN.
For the same reasons as in case (c) in section \ref{sec5_3b}, the abundances $n$/H, D/H, $^7$Li/H, and $^6$Li/H are higher while $^3$He/H, and $^7$Be/H are lower than in SBBN.


\section{Summary}\label{sec6}
We studied effects of the $f(G)$ gravity on BBN.  It was assumed that a $f(G)$ term exists in the BBN epoch of $T_9=[100,~0.01]$.  The functional form was taken to be $\xi =df/dG =\xi_0 (t/t_0)^p$ where $t_0=1$ s is a typical time scale, and $\xi_0$ and $p$ are parameters.  Under this assumption, the $f(G)$ model can be described with three parameters, i.e., the coefficient $\xi_0$, the power-law index $p$, and the initial value of $f(G_{\rm i})$.  We then showed a method to solve physical variables during BBN consistently taking account of the modified cosmic expansion rate.

Numerical calculations of BBN in the $f(G)$ model were performed, and results were analyzed.  It was then found that a proper solution for the cosmic expansion rate during the BBN epoch does not exist in some parameter region.  In addition, we compared calculated results of primordial light element abundances with observational data.  Constraints on parameters of the $f(G)$ gravity are then derived from the observed abundances.  Since a change in the cosmic expansion rate easily deviates primordial D abundance from observational limits, the observed D abundance gives the strongest constraint on the parameters (figures \ref{fig2}, \ref{fig3}, \ref{fig4}).  We thus obtained allowed parameter regions where the universe expands properly and final abundances of light elements in the BBN epoch are consistent with observations.

In the case of $\xi_0 >0$ and $f(G_{\rm i})=0$ (section \ref{sec5_1}), the existence of the solution for the cosmic expansion rate determines the allowed parameter region of $\xi_0$ and $p$ predominantly. The constraint is derived as
\begin{eqnarray}
\label{eq40}
\log \xi_0 \lesssim \left\{
\begin{array}{ll}
  71 -\frac{11}{6} \left( 3.5-p \right) & ({\rm for}~p \lesssim 3.5) \\
  \left(71-70\right) & ({\rm for}~3.5 \lesssim p \lesssim 4.2) \\
  70 -6\left( p-4.2 \right) & ({\rm for}~4.2 \lesssim p).  \\
\end{array} \right.
\end{eqnarray}
In the case of $\xi_0 <0$ and $f(G_{\rm i})=0$ (section \ref{sec5_2}), on the other hand, the parameter region is constrained from the limit on elemental abundances for $p<3.8$ and from the existence of the solution for $3.8 <p$. The derived constraint is
\begin{eqnarray}
\label{eq41}
\log \left|\xi_0 \right|\lesssim \left\{
\begin{array}{ll}
  74 -\frac{7}{4} \left( 3.8-p \right) & ({\rm for}~p \lesssim 3.8) \\
  74 -12.5 \left( p -3.8 \right) & ({\rm for}~3.8 \lesssim p \lesssim 4) \\
  71.5 -6.5\left( p-4 \right) & ({\rm for}~4 \lesssim p). \\
\end{array} \right.
\end{eqnarray}
In the case of $\xi_0 >0$ and $p=2$ (section \ref{sec5_3}), we derived a constraint on the parameter region of $\xi_0$ and $[f(G)/(\xi G)]_\mathrm{i}$ which is the ratio of the quantities $f(G)$ and $(\xi G)$ at the initial time of computation.  The constraint comes from the existence of the solution for $[f(G)/(\xi G)]_\mathrm{i}<2$, and from the limit on elemental abundances for $2<[f(G)/(\xi G)]_\mathrm{i}$.  The derived constraint is
\begin{eqnarray}
\label{eq42}
\log \xi_0 \lesssim \left\{
\begin{array}{ll}
  68 & ({\rm for}~[f(G)/(\xi G)]_\mathrm{i} \lesssim 2) \\
  71 & ({\rm for}~2 \lesssim [f(G)/(\xi G)]_\mathrm{i}).  \\
\end{array} \right.
\end{eqnarray}
In the case of $\xi_0 <0$ and $p=2$ (section \ref{sec5_4}), in contrast, the parameter region is constrained from the limit on elemental abundances for $[f(G)/(\xi G)]_\mathrm{i}<2$ and from the existence of the solution for $2 <[f(G)/(\xi G)]_\mathrm{i}$. The derived constraint is
\begin{eqnarray}
\label{eq43}
\log \left| \xi_0 \right| \lesssim \left\{
\begin{array}{ll}
  71 & ({\rm for}~[f(G)/(\xi G)]_\mathrm{i} \lesssim 2) \\
  68 & ({\rm for}~2 \lesssim [f(G)/(\xi G)]_\mathrm{i}).  \\
\end{array} \right.
\end{eqnarray}
These are constraints on the $f(G)$ term during the BBN epoch.

\acknowledgments
This work was supported in part by the National Research Foundation of Korea (NRF) (Grant Nos. NRF-2015R1A2A2A01004727 and NRF-2014R1A2A2A05003548).  S.K. was supported by Basic Science Research Program through the NRF funded by the Ministry
of Education (NRF-2014R1A1A2059080).



\end{document}